\documentclass[letter,11pt]{article}
\usepackage{setspace}
\usepackage[english]{babel}
\usepackage[latin1]{inputenc}
\usepackage[pdftex]{graphicx}
\usepackage{theorem}
\usepackage{amsmath}
\usepackage{ifthen}
\usepackage{commath}
\usepackage{amsfonts}
\usepackage{float}
\usepackage{color}
\usepackage{bibentry}
\usepackage{natbib}
\usepackage{hyperref}
\usepackage[all]{xy}
\usepackage{epstopdf}
\usepackage{caption}
\usepackage{subcaption}
\usepackage{colortbl}
\usepackage{array,booktabs,tabularx}
\usepackage{url}
\usepackage{amssymb}

\newcommand{\YObs}{Y_{{\rm Obs}}}
\newcommand{\Ynwp}{Y_{{\rm NWP}}}
\newcommand{\yobs}{y_{{\rm Obs}}}
\newcommand{\ynwp}{y_{{\rm NWP}}}

\newcommand{\cov}{\rm cov}
\newcommand{\s}{{\rm S}}

\newcommand{\dd}{{\rm d}}
\newcommand{\ANLDataText}[1]{}
\newcommand{\PropYbase}{\Psi} 

\begin{document}
\title{Stochastic simulation of predictive space-time scenarios of wind speed using observations and physical model outputs}
\author{Julie Bessac \textsuperscript{1}, Emil Constantinescu \textsuperscript{1,2}, Mihai Anitescu \textsuperscript{1,2} \thanks{\bf Preprint ANL/MCS-P5432-1015}}
\date{}
\vspace{20mm}
\maketitle

{\small $^1$ \em{Mathematics and Computer Science Division, Argonne National Laboratory, Argonne, IL}}

{\small $^2$ \em{The University of Chicago, Chicago, IL}}

\begin{abstract}
We propose a statistical space-time model for
predicting atmospheric wind speed based on deterministic numerical weather
predictions and historical measurements. We consider a Gaussian
multivariate space-time framework that combines multiple sources of
past physical model outputs and measurements in order to produce a probabilistic wind speed
forecast within the prediction window. We illustrate this strategy on wind speed
  forecasts during several months in 2012 for a region near the Great Lakes
  in the United States. The results show that the prediction is improved in
  the mean-squared sense relative to the numerical forecasts as well as in probabilistic
  scores. Moreover, the samples are  shown to produce
  realistic wind scenarios based on sample spectra and space-time correlation structure. 
\end{abstract}

\section{Introduction}\label{SecIntro}

In this study we propose a statistical space-time model for predicting
atmospheric wind speed based on numerical weather predictions and
historical measurements.  
The wind speed predictions are based on
deterministic numerical weather prediction (NWP) model outputs in a
framework that integrates past dependence between observational
measurements and the NWP model outputs. 
The past dependence between these two datasets is modeled linearly
into a Gaussian process (GP) hierarchical framework. The aim of this
work is to
  improve the wind speed forecasts and to produce samples (referred to
  as scenarios) from the predictive distribution of wind speeds. This
  is achieved by using GP framework in conjunction with NWP forecast
  output.  

  Atmospheric near-surface wind conditions are important for numerous sectors of human activities, and the topic has received considerable attention in the past several years, for instance, in the study of crop models \cite{Brisson03}, object drift into the ocean \cite{Ailliot06drift}, and severe weather forecasting \cite{Thorarinsdottir12}. 
Arguably, one of the largest applications is in wind energy, because
energy management with renewables relies heavily on an accurate description of the forecast uncertainty \cite{Pinson09,Constantinescu11,Pinson13,Papavasiliou15,Li15}.

\subsection{Existing context on statistical wind speed prediction}

Several components of the wind field can be predicted separately or jointly: the zonal and meridional components \cite{Hering10,Sloughter13}, wind speed \cite{Brown84,Gneiting06,Sloughter10}, and wind direction \cite{Bao10}.  
Prediction of wind conditions can be generated by statistical models built for predicting observed wind \cite{Brown84,Gneiting06,Hering10}, or they can be generated from the statistical post-processing of NWP model forecasts; these latter fall into the domain of model output statistics (MOS) methods \cite{Glahn72,Raftery05,Gneiting05}.  
From a statistical point of view, the prediction error can be accounted for through the use of predictive distribution. 
Ensemble forecasts aim at assessing the uncertainty associated with the numerical model; however, this strategy is known to be often uncalibrated and underdispersive \cite{Gneiting05}.  
More recently, the generation of predictive scenarios has gained significant momentum. 
These scenarios enable accounting for the uncertainty of the forecasts for various locations and$\slash$or time-ahead lags \cite{Pinson08,Pinson12}.  

%
%

In the context of improving numerical forecasts, MOS methods provide probabilistic forecasts by post-processing the single or ensemble forecasts and tend to address the issue of bias and dispersion.  
Introduced at first for single-trajectory forecast \cite{Glahn72}, MOS methods have been extended to post-processing methods EMOS (ensemble model outputs statistics), for ensemble forecasts \cite{Gneiting05}. 
Most of the MOS methods for ensembles are variations and extensions of the Bayesian model averaging (BMA) initially proposed in \cite{Raftery05} and the non-homogeneous regression (NGR) model proposed in \cite{Gneiting05}. 
Several variants of both models have been proposed with different distributions \cite{Sloughter10,Lerch13,Baran14a,Baran14b}, a BMA extension to spatial interpolation (see \cite{Scheuerer15} and references therein), and an NGR with regime-switching \cite{Baran15}.  
Lately, multivariate models have been introduced, namely, spatial models \cite{Thorarinsdottir10,Gel04} and bivariate frameworks \cite{Sloughter10,Schuhen12}.  
In \cite{Schefzik13} a multivariate tool based on the use of copulas is presented that allows one to account for time-ahead dependence, spatial dependence and dependence among variables. 

%
%

MOS and regression methods treat the NWP model outputs as covariates. 
Following the arguments pointed out in \cite{Schefzik13}, these models neglect the dependence between variables, between the different time-ahead lags, and between spatial locations. 
Notable exceptions are the use of copulas to model the variable dependence \cite{Schefzik13} and the use of a parametric model for spatial locations \cite{Feldmann14}. 
Moreover, the use of covariates creates difficulties such as addressing mis-alignment in space and$\slash$or time with the response variables and possible non-exhaustive sampling of these covariates.  
Multivariate modeling alleviates these problems in modeling jointly several variables as a random process.  
Multivariate space-time modeling has been an area of intense research in the past two decades; see \cite{Fanshawe12} for a review of bivariate geostatistical modeling, and see \cite{Berliner00}, for a discussion of hierarchical Bayesian modeling for multiple dependent datasets.   

%
%

In the multivariate modeling context, various statistical approaches have been proposed for hybrid NWP -- physical observations usage.  
For example, in \cite{Fuentes05} a Bayesian hierarchical model is presented that combines NWP outputs and observed measurements to provide spatial prediction for chemical species.  
A hidden process is used to represent the unobserved ``true'' concentration of sulfur dioxide, and the sources of data are affine transformations of this ``true'' process. 
A similar approach was used in a space-time context for multiple measurements of snow water equivalent data in  \cite{Cowles02}. 
Several outputs of regional climate models are combined in a spatial framework by using a hierarchical model based on a spatial random effects model in \cite{Kang12}. 
Berrocal et al. \cite{Berrocal12} proposed a space-time hierarchical Bayesian model to fuse measurements and numerical model outputs of air-quality data, with an extension of a downscaling model introduced some years ago.  
Royle and Berliner \cite{Royle99a} introduced a conditional hierarchical model that combines two heterogeneous spatial datasets of ozone and temperature in the objective to predict these two variables on gridded points while they are recorded at two different irregular networks.  
A joint distribution is used to cope with the non-alignment of the two datasets and of the predicted values at gridded points.  
This distribution is specified in a hierarchical conditional way in order to avoid the direct specification of the joint distribution.  
Indeed, the modeling of multivariate covariance structure is challenging and is still an on-going research area; see \cite{Apanasovich10,Genton14,Bourotte15}. 

%
%

\subsection{Proposed modeling framework and position within the existing literature} 

In this study we fuse two heterogeneous datasets: numerical forecasts and physical observations. 
Our endeavor stems from the observation that physical observations alone cannot be used to deliver accurate forecasts 24-48-hours ahead, whereas NWP uncertainty analysis may be (as pointed out above) uncalibrated and underdispersive.
By heterogeneity, we mean that the physical observations are not necessarily on the NWP output grid. 
We note that refining the grid would be a costly computational expense
and would not guarantee that the sites of the physical observations were
exactly on the grid or that the forecast were improved. 
Moreover, NWP physics constrains assumptions that make refining below a grid size limit inconsistent, if not even erroneous \cite{Palmer14}.  
We propose a bivariate space-time Gaussian process to improve forecasts from an NWP model, where the physics model outputs and the measurements are modeled as two random space-time processes. 
GPs allow us to describe the conditional forecast distribution as a Gaussian distribution as well as facilitate robust computational algorithms \cite{Anitescu12,Stein12b}. 
Moreover, we use a marginal Box-Cox transformation to address
 the typical positivity and skewness of wind speed distribution. 
Our model is specified in a hierarchical way in order to avoid characterizing of the full space-time bivariate covariance.    
We extend the specification, which was initially proposed in \cite{Royle99a,Royle99b} in a spatial context, to space-time modeling.

 The numerical forecasts of wind speed are combined with historical measurements data to provide scenarios of prediction. 
 The modeling of the NWP forecasts as a random process with spatial or space-time structure, also carried out in \cite{Fuentes05,Berrocal12,Royle99b}, differentiates this approach from MOS methods, where dependencies are most of the time ignored.  
 The NWP random process modeling allows for a consistent statistical framework for the discrepancy between the NWP output and physical measurement locations. 
A particularly important aspect of our model is that the proposed prediction framework accounts for the space-time dependence between these two heterogeneous datasets. 
Furthermore, in this work we consider a strategy where NWP
  future and past temporal information is used to provide a temporal
  prediction that begins at the current time.  
  In our approach we set a 24-hour ahead forecast window and use this
  entire window of NWP outputs to predict wind speed at each time during
  this window. In contrast, MOS methods commonly work with current
  information only. In other words, MOS methods adjust the forecast
  point-wise in space or time sequentially one step at a time;
  whereas, our approach accounts for future and past information
  within the forecast window itself. 
The same argument applies to the strategies proposed in \cite{Fuentes05b,Berrocal12}.  

%
%

The paper is organized as follows. 
In Section \ref{SecModel} we introduce the modeling context and the
statistical formulation.  
In Section \ref{SecData} we describe the two sources of data that are used and combined. 
In Section \ref{SecRes} the model is validated on different months of the year, and the quality of the space-time prediction at one out-of-sample station is assessed. 
We highlight the improvements in terms of the forecasting accuracy of the proposed model with respect to the NWP forecasts. 
We conclude in Section \ref{SecConclu} by presenting general improvements made by the model with respect to the NWP data.


\section{Statistical model for NWP outputs}\label{SecModel}

In this section we introduce a Gaussian modeling framework that embeds the space-time dependence between measured observations and NWP model forecasts.   
We model two heterogeneous spatio-temporal datasets as jointly distributed variables. 
We extend the hierarchical GP ideas of \cite{Royle99a} to a space-time context; the joint process is a space-time Gaussian process specified conditionally.  
We provide temporal prediction of observations given past and future NWP data while accounting for past space-time dependencies between the two datasets; see Equation \eqref{EqSetup} below. 

\subsection{Overview of the proposed method} 
We consider measured observations, $\YObs$, and numerical model (NWP)
forecasts, $\Ynwp$. Both are available in the past, however, only
future $\Ynwp$ is available at the current time. We are therefore
interested in generating samples from the distribution of future observations
conditional on the past observations and NWP simulations and on future
NWP simulations. This distribution  
is represented as a hierarchical Gaussian process and is calibrated by
maximizing its likelihood. In particular, the ingredients of our
proposed approach are given as follow.
\begin{itemize}
  \item[-] We aim to construct a probabilistic model for future observations based
on the current available data:
\begin{align}
   \label{eq:full:pred:distr:abstract}
&  p\left(\yobs^{u}(t_{\mathrm{future}})\, |\, \theta , \, \ynwp^{a}(t_{\mathrm{future}}),\,
  \yobs^{a}(t_{\mathrm{past}}),\, \ynwp^{a}(t_{\mathrm{past}})\right)\,,
\end{align}
where the superscript ``$a$''  stands for available and  ``$u$'' for
unavailable quantities and $\theta$ denotes parameters of a
statistical model (see Section \ref{sec:model:context}). 
\item[-] To specify the probability in
  \eqref{eq:full:pred:distr:abstract}, we use a joint model for $(\YObs,\Ynwp)$ represented by a GP. As indicated in
  Section \ref{sec:Hier:Biv:Model}, the full joint distribution can be
  expressed as
\begin{align}
 \label{eq:GP:full}
 \begin{pmatrix}\YObs \\ \Ynwp  \end{pmatrix} \sim \mathcal{N}\left( \begin{pmatrix} \mu_{\rm{Obs}}(\theta) \\ \mu_{\rm{NWP}}(\theta) \end{pmatrix}  , \begin{pmatrix} \Sigma_{\rm{Obs}}(\theta)  & \Sigma_{\rm{Obs},\rm{NWP}}(\theta) \\   \Sigma_{\rm{Obs},\rm{NWP}}^{T}(\theta)& \Sigma_{\rm{NWP}}(\theta)
\end{pmatrix}\right), 
\end{align}
where $\theta$ is the set of parameters that describe the parametric
shapes of the means and covariances.
\item[-] The joint distribution of $(\YObs,\Ynwp)$ in \eqref{eq:GP:full} is specified in a
  conditional hierarchical way where the distributions of
  $(\YObs|\Ynwp)$ and $(\Ynwp)$ are described separately in terms of
  parametric space-time mean and covariance. This is discussed in
  detail in Section \ref{sec:Stat:Model}. 
\item[-] The parameters $\theta$ in \eqref{eq:GP:full} are estimated
  by maximizing the 
  likelihood, as discussed in Section \ref{sec:estimPar}, so that 
$\theta^* =  \arg\max_\theta~ \mathcal{L} \left(\theta
; \allowbreak\yobs^{a}(t_{\mathrm{past}}),
  \allowbreak\ynwp^{a}(t_{\mathrm{past}})\right)$. This aspect is also 
   discussed at the end of Section \ref{sec:model:context} in the more
   abstract context of Equation \eqref{eq:full:pred:distr:abstract} as well
   as specify 
    the probabilistic modeling assumptions.     
\item[-] The predictive scenarios sampled from $p\left(\yobs^{u}(t_{\mathrm{future}})|\cdots\right)$  are obtained via
  kriging equations and 
  are detailed in Section \ref{sec:kriging}. 
\end{itemize}
In other words, we represent the joint distribution of the numerical
simulations and observations through a Gaussian model that is
calibrated by maximizing the likelihood of the model parameters. This
distribution is then used to condition on the numerical forecasts at the current time
in order to forecast observations. In the following sections we give
details of each step in constructing this framework. 

\subsection{Modeling context\label{sec:model:context}}
Let us assume that both measured observations $\YObs$ and NWP forecasts $\Ynwp$ are available from time $t_{1}$ to time $t_{k_{K}}$.  
In the following, the term ``observations'' refers to the observational measurements. 
Observations are available at $J_{0}$ locations $\s=\{s_{1},...,s_{J_{0}}\}$, and NWP forecasts are available over a grid that covers these stations.  
Each day, the Weather Research and Forecasting (WRF) model is run for a period of $h$ hours independently from the
previous day, because WRF is initialized from a reanalysis or
assimilated dataset; time can be written in terms of blocks of length $h$.  
Henceforth, we consider a time window of $h=24$ hours. 
We denote by $b_i$ the $i$th time block of length $h$, $b_i=\{t_{k_{i}},...,t_{k_{i}+h-1}\}$.

The objective here is to predict the measurements $\YObs$ between time $t_{k_{K+1}}$ and $t_{k_{K+1}+h-1}$ at stations $\s=\{s_{1},...,s_{J_{0}}\}$, and possibly at locations $\rm{S}_{0}=\{s_{J_{0}+1},...,s_{J}\}$ where no historical measurements are recorded, from NWP forecasts  that are available between $t_{k_{K+1}}$ and $t_{k_{K+1}+h-1}$.  
This can be summarized by
\begin{align}
\label{EqSetup} 
\left( \begin{array}{c}
\yobs^{a}(b_{1:K};\s) \\
\yobs^{u}(b_{K+1};\s,\rm{S}_{0})
\end{array} \right) \textrm{ and } 
\left( \begin{array}{c}
\ynwp^{a}(b_{1:K};\s) \\
\ynwp^{a}(b_{K+1};\s,\rm{S}_{0})
\end{array} \right). 
\end{align}

In this context the model is trained on the following available pairs: 
\begin{align}
\nonumber& \Big(\yobs^{a}(b_{1:K};\s) , \ynwp^{a}(b_{1:K};\s)\Big), 
\end{align}
 and the prediction is made from
 $\ynwp^{a}(b_{K+1};\s,\rm{S}_{0})$ to estimate
 $\allowbreak\yobs^{u}(b_{K+1}; \allowbreak \s, \allowbreak \rm{S}_{0})$, where
 $b_{K+1}=\{t_{k_{K+1}},..., t_{k_{K+1}+h-1} \}$.
In a probabilistic sense, we aim to compute
\begin{align}
   \label{eq:full:pred:distr}
&  p\left(\yobs^{u}(b_{K+1}) | \ynwp^{a}(b_{K+1}),
  \yobs^{a}(b_{1:K}), \ynwp^{a}(b_{1:K})\right) = \\
  \nonumber
& \quad\int p\left(\yobs^{u}(b_{K+1}),\theta | \ynwp^{a}(b_{K+1}),
   \yobs^{a}(b_{1:K}), \ynwp^{a}(b_{1:K})\right) \: \dd \theta\,, 
\end{align}
where $\theta$ is a random set of model parameters, blocks $b_{1:K}$
are available, and $b_{K+1}$ is a predicted block; the spatial
components are suppressed for brevity. 
Note that $b_{K+1}$ is not
necessarily a block coming right after $b_{K}$, but rather a day that
is not observed. 
To simplify the computation of
\eqref{eq:full:pred:distr},
we now make several assumptions. 
First, we assume ($i$)
that we have approximate independence of $\yobs^u(b_{K+1})$ on
$\yobs^{a}(b_{1:K}), \ynwp^{a}(b_{1:K})$
conditional on $\ynwp^{a}(b_{K+1})$. 
In hierarchical models such as ours, which has NWP predictions as its first
layer and the observation sites as the second layer, one
commonly assumes that random variables on the second layer
are independent conditional on the realizations of the ones
in the first layer; see \cite{Cressie11}.  
This assumption is correct if the
additional randomness occurs from the noise of different,
unrelated sensors. 
In our case, since we are considering the
error of NWP models, the difference between prediction
and observations most likely is due to features not
modeled by NWP. 
They may be the use of lower resolution
or models that have been obtained by some level of
space-time homogenization of the physics of the model considered. 
In this case, the difference is the modeling of subscale
noise, which can be assumed to have short temporal
correlation scales; see \cite{Majda06}. 
Moreover, under a short temporal correlation of error assumption, 
our use of 24-hour temporal blocks as opposed to an every time-index would 
strengthen the validity of approximate conditional
independence on NWP simulations of wind realizations at
observation sites. The independence of 
$\yobs^u(b_{K+1})$ on the $\ynwp^{a}(b_{1:K})$
 conditional on $\ynwp^{a}(b_{K+1})$ may also be a good
 approximation given the short temporal correlation scales
 of subscale noise discussed above. 

As a result, assumption ($i$) implies that 
 the integrand in
\eqref{eq:full:pred:distr} can be approximated as 
\begin{align}
  \nonumber
  &p\left(\yobs^{u}(b_{K+1}), \theta | \ynwp^{a}(b_{K+1}),
  \yobs^{a}(b_{1:K}), \ynwp^{a}(b_{1:K})\right)
\\
\nonumber
&\quad\approx p\left(\yobs^{u}(b_{K+1}) |
\theta,\ynwp^{a}(b_{K+1}), \yobs^{a}(b_{1:K}), \ynwp^{a}(b_{1:K})\right)
\\
\nonumber
&\qquad\qquad p\left(\theta |
  \yobs^{a}(b_{1:K}),\ynwp^{a}(b_{1:K})\right) 
  \\
  & \quad\approx p\left(\yobs^{u}(b_{K+1}) |
  \theta,\ynwp^{a}(b_{K+1}) \right) 
  p\left(\theta|
  \yobs^{a}(b_{1:K}),\ynwp^{a}(b_{1:K})\right)\,.
      \label{eq:full:distr0}
\end{align}
The first approximation above comes from the fact that
  conditioning $\theta$ on $\ynwp^{a}(b_{K+1})$ will not alter the
  equation by a measurable amount given large $K$.  

In this study, we assume ($ii$) that $\theta^*$ can be obtained by maximizing the likelihood
\begin{align}
\label{eq:abstract:likelihood}
\theta^* &= {\rm argmax}_\theta~ \mathcal{L}\left(\theta
;\yobs^{a}(b_{1:K}),\ynwp^{a}(b_{1:K})\right) \\
\nonumber
&= {\rm argmax}_\theta~   p\left(\theta |
  \yobs^{a}(b_{1:K}),\ynwp^{a}(b_{1:K})\right)\, 
\end{align}
With assumptions ($i$-$ii$) 
and thus using \eqref{eq:full:distr0} in
\eqref{eq:full:pred:distr}, we obtain that
\begin{align} 
  \label{eq:full:distr}
  & \int p\left(\yobs^{u}(b_{K+1}),\theta | \ynwp^{a}(b_{K+1}),
  \yobs^{a}(b_{1}),...,\yobs^{a}(b_{K}),
  \right.\\
    &\nonumber
    \quad \left. 
 \ynwp^{a}(b_{1}),...,\ynwp^{a}(b_{K})\right) \: \dd \theta \approx 
 p\left(\yobs^{u}(b_{K+1})| \theta^*,\ynwp^{a}(b_{K+1})
\right). 
\end{align}
This argument holds well if $\theta^{*}$ has a
narrow distribution. This is indeed our situation as illustrated
in Fig. \ref{fig:parameteruncert}. 
In what follows we consider multivariate normal
distributions for \eqref{eq:full:distr}.
We have found that a sensible approach is to model statistically the
output of NWP itself. 
In other words, one can consider
NWP as a noisy realization of a latent underlying process NWP$^V$
(which models the evolution of spatially averaged
quantities). 
With the NWP conditional on this NWP$^V$ we then assume it
to be independent for two different temporal blocks; that is, all
temporal correlation between successive blocks is due to NWP$^V$ itself. 
Note that we never forecast the NWP output
using the statistical model we develop; we forecast only its
relationship to the observations. 
Thus, we do not need to model
explicitly the temporal correlation between different blocks
of NWP as long as a sample is produced by the WRF model
by a 
black-box 
mechanism that emulates the correct interblock correlation
by its relationship to NWP$^V$. 
Moreover, if such an assumption does
not hold completely, it can lead only to more conservative
forecasts.

To summarize our approach, for a given statistical model, we first estimate
$\theta^*$ from the available data (model and observations)
using \eqref{eq:abstract:likelihood}. Then, using
\eqref{eq:full:distr}, we obtain a predictive
distribution by conditioning only on the NWP
predictions for the same temporal block
and plugging-in the maximum likelihood
estimate $\theta^*$: 
\begin{align}
  \nonumber
  &p\left(\yobs^{u}(b_{K+1}) | \ynwp^{a}(b_{K+1}),
  \yobs^{a}(b_{1:K}), \ynwp^{a}(b_{1:K})\right) \\
  &  \label{eq:pred:distr}\qquad \approx  p\left(\yobs^{u}(b_{K+1}) |
  \theta^*,\ynwp^{a}(b_{K+1}) \right)\,. 
\end{align}
These
choices are  motivated by computational tractability; by
the fact that we assume that the information missed by NWP
is subscale-type information, which, as mentioned above,
is assumed to have short time correlations conditional on
NWP realizations, by our assumption that the likelihood is
  narrow around the optimum $\theta^{*}$ (see
Fig. \ref{fig:parameteruncert}) and by the fact that we do not
forecast 
NWP itself but rather the relationship between NWP and observations.  
In Section
\ref{sec:Hier:Biv:Model} we review a hierarchical approach for
Gaussian processes, and in Section \ref{sec:Stat:Model} we present the model
used for the mean and covariance functions that introduce the
parametrization $\theta$. 

\subsection{Hierarchical bivariate framework\label{sec:Hier:Biv:Model}}
Gaussian processes are chosen for their convenience in expressing
conditional distributions and in a multivariate and space-time context. 
Power transformations are commonly used to approximate Gaussian margins.  
To address the typical skewness of the wind speed
distribution, we apply the Box-Cox transformation; see, for instance, \citep*{Brown84}.  
A specific transformation is used for each dataset (NWP and measurements) to account for the heterogeneity between the two datasets.  
Within each dataset, the same power transformation is applied to each spatial location to preserve the variance structure. 
The model is fitted on the transformed data.  
We write the joint distribution of the process $(\YObs,\Ynwp)$ as 
\begin{align}
 \label{fulldistri0}
 \begin{pmatrix}\YObs \\ \Ynwp  \end{pmatrix} \sim \mathcal{N}\left( \begin{pmatrix} \mu_{\rm{Obs}} \\ \mu_{\rm{NWP}} \end{pmatrix}  , \begin{pmatrix} \Sigma_{\rm{Obs}}  & \Sigma_{\rm{Obs},\rm{NWP}} \\   \Sigma_{\rm{Obs},\rm{NWP}}^{T} & \Sigma_{\rm{NWP}} 
\end{pmatrix}\right).
\end{align}
The positive-definiteness of block matrices is generally difficult to
ensure when specifying the three blocks in \eqref{fulldistri0} independently.  
Therefore, to avoid specifying of the full covariance in \eqref{fulldistri0}, we follow the hierarchical conditional modeling proposed by \cite{Royle99a,Royle99b}, and we model  $(\YObs|\Ynwp)$ and $(\Ynwp)$,  where $(\YObs|\Ynwp)$ stands for the conditional distribution of $\YObs$ given $\Ynwp$. 
When $(\YObs,\Ynwp)$ is a Gaussian process, $(\YObs|\Ynwp)$ and $(\Ynwp)$ follow a Gaussian distribution; then only first- and second-order structures are to be specified.  
Consequently the model is described by the following distributions: 
\begin{align}\label{districond}
\left( \YObs | \Ynwp  \right) & \sim \mathcal{N}\Big( \mu_{Obs|NWP}, \Sigma_{Obs|NWP} \Big). 
\end{align}
The Gaussian joint distribution of  $(\YObs,\Ynwp)$ implies a conditional linear dependence between  $\YObs$ and $\Ynwp$, which agrees reasonably with the data analysis: 
\begin{align}\label{meancond}
& \mu_{Obs|NWP} = \rm{E}(\YObs|\Ynwp) = \mu + \Lambda \Ynwp, 
\end{align}
where $\mu$ and $\Lambda$, which is called the transition matrix, will be parameterized below, and 
\begin{align}\label{distriNWP}
\Ynwp  & \sim \mathcal{N}\Big( \mu_{NWP}, \Sigma_{NWP} \Big).
\end{align}

From these equations, we express the full joint distribution given by \eqref{fulldistri0} as    
\begin{align}
\label{fulldistri}
\begin{pmatrix}\YObs \\ \Ynwp  \end{pmatrix} \sim \mathcal{N}\left( \begin{pmatrix} \mu + \Lambda \mu_{NWP}\\ \mu_{NWP} \end{pmatrix}  , \begin{pmatrix} \Sigma_{Obs|NWP} + 
\Lambda  \Sigma_{NWP} \Lambda^{T}  & \Lambda \Sigma_{NWP} \\  (\Lambda \Sigma_{NWP})^{T} & \Sigma_{NWP} 
\end{pmatrix}\right).
\end{align}

\subsection{Statistical model\label{sec:Stat:Model}}
To provide time prediction and to ensure model
parsimony, we propose a parameterization in space and time of the
involved quantities such that the first- and second-order structures of
the conditional and the marginal distributions defined by
\eqref{districond} and \eqref{distriNWP} are specified following an exploratory analysis of the datasets.

\subsubsection{Marginal mean structure of $(\Ynwp)$}
The empirical mean function of $\Ynwp$  exhibits spatial patterns associated with the geographical coordinates but also with several parameters of the NWP model. 
Indeed the studied area is in the Great Lakes region with the large water mass of Lake Michigan (the ``land use'', LU, is used, which is a categorical variable that represents the type of land used in the parameterization of the NWP model). 
Time-periodic effects are present in the first-order structure of $\Ynwp$ and are accounted for through harmonics of different frequencies. 
In Figure \ref{fig:mean_hour}, these spatial and temporal patterns are plotted. 
We write
\begin{align}
  \label{eq:EqmuNWP}
\mu_{NWP}(t,s) = \rm{E}(\Ynwp(t,s)) &= \left(\beta_0 + \beta_1 \cos\left(\frac{2\pi
  t}{24}\right) + \beta_2 \sin\left(\frac{2\pi t}{24}\right) + \beta_3
\cos\left(\frac{2\pi t}{12}\right)  + \right. \\
\nonumber & \qquad \left.  \beta_4 \sin\left(\frac{2\pi
  t}{12}\right) +\beta_5 \cos\left(\frac{2\pi t}{8}\right) + \beta_6
\sin\left(\frac{2\pi t}{8}\right)\right) \\ 
\nonumber &\qquad \left(\alpha_{0}^{(\rm{LU}(s))} + \alpha_{1}(s)\right),
\end{align}
where $t$ is measured in hours, $\rm{LU}(s)$ is an integer that represents the land use associated with station $s$ used in the model;  $(\alpha_{0}^{(l)})_{l=1,...,n}$, with $n$ the number of possible land uses, $(\alpha_{1}(j))_{j=1,...,J_{0}}$ and $(\beta_{k})_{k=0,...,6}$ are real numbers to be estimated. 

\subsubsection{Marginal covariance structure of $(\Ynwp)$}\label{sec:covNWP}
The block structure of the space-time covariance of the data suggests expressing
wind speed at each station as a linear transformation
of an unobserved common signal $\rm{Y}_{0}$ with added noise; see Figure \ref{fig:map_corr}.  
Intuitively we can think of this common signal as an average flow over
the studied region. The wind speed at each site is a linear transformation of this average flow. 
The temporal dynamics of the unobserved signal is modeled with a squared exponential covariance. 
The following structure is used: 
\begin{align}
\nonumber \rm{Y}(b_{i},s_{j}) = \PropYbase_{s_{j}} Y_{0}(b_{i}) + \varepsilon_{s_{j}}(b_{i}),
\end{align}
where $\rm{Y}$ represents $\Ynwp$ in this paragraph and will represent $(\YObs|\Ynwp)$ in Section \ref{sec:condVar}, $b_{i}$ is a temporal window of $h=24$ lags, $s_j$ the spatial location, and $\PropYbase_{s_j}$ is an $h \times h$-matrix. 
The various $\varepsilon_{s_j}$ are assumed independent from each other and from $Y_{0}$.  
This model is inspired in part by an earlier study \citep{Constantinescu13} where the $\PropYbase$ operators were used to
represent a known functional relation. 
In our case, $\PropYbase_{s_j}$ is a parameterized matrix that is inferred from the data. 

The overall space-time covariance of $\Ynwp$ has the following structure:
\begin{align}\label{eq:EqCov}
\Sigma_{NWP}(.,s_{i};.,s_{j}) = \cov(\Ynwp(.,s_{i}),\Ynwp(.,s_{j})) = (\PropYbase_{s_{i}} \Gamma_{0}  \PropYbase_{s_{j}}^{T}) + \delta_{i-j} \Gamma_{s_i},
\end{align}
for $j \in \{1,...,J_{0}\}$ and where $\delta$ stands for the Kronecker symbol.  
The $h \times h$-matrices $\Gamma_{s_j}$ are written as
\begin{align}
\nonumber \Gamma_{s_j}[l,k] = \sigma_{s_j} \exp(- \lambda_{s_j} (|t_k - t_l|)^{2}) +  \delta_{k-l} \gamma_{s_j}
\end{align}
and 
\begin{align}
\nonumber \Gamma_{0}[l,k] = \sigma_{0} \exp(- \lambda_{0} (|t_k - t_l|)^{2}) +  \delta_{k-l} \gamma_{0}, 
\end{align}
 where $j \in \{1,...,J_{0}\}$, $\sigma_{s_j}$, $\lambda_{s_j} $ and $\gamma_{s_j}$, and $\sigma_{0}$, $\lambda_{0}$, and $\gamma_{0}$ are positive real numbers to be estimated. 

Following the data analysis, the $h\times h$-matrices $\PropYbase_{s_j}$ are parameterized as tridiagonal matrices. 
Given the study of the variance in space and time, the diagonal and off-diagonal quantities  are modeled with a quadratic dependence in time and spatially dependent coefficients.  
The diagonal, subdiagonal, and superdiagonal of the matrix $\PropYbase_{s_j}$ are written respectively as 
\begin{align}
  \nonumber  \PropYbase_{s_j}[i,i] &= \left(1 + \nu_{1}{\rm Lat}(s_j) + \nu_{2}{\rm  Long}(s_j)\right) +  \\
\nonumber & \qquad \left(1 + \nu_{3}{\rm Lat}(s_j) + \nu_{4}{\rm  Long}(s_j)\right) \times i  + \\
\nonumber & \qquad \left(1 + \nu_{5}{\rm Lat}(s_j) + \nu_{6}{\rm  Long}(s_j)\right) \times i^{2}\,,\\
\nonumber \PropYbase_{s_j}[i,i-1]&= (1 + \nu_{7}{\rm Lat}(s_j) + \nu_{8}{\rm  Long}(s_j)) +  \\
\nonumber & \qquad(1 + \nu_{9}{\rm Lat}(s_j) + \nu_{10}{\rm Long}(s_j)) \times i  +   \\
\nonumber  & \qquad (1 + \nu_{11}{\rm Lat}(s_j) + \nu_{12}{\rm Long}(s_j)) \times i^{2}, \\
\nonumber \PropYbase_{s_j}[i,i+1] &= (1 + \nu_{13}{\rm Lat}(s_j) + \nu_{14}{\rm  Long}(s_j)) +  \\
\nonumber & \qquad(1 +  \nu_{15}{\rm Lat}(s_j) + \nu_{16}{\rm Long}(s_j)) \times i  +  \\
\nonumber  & \qquad  (1 + \nu_{17}{\rm Lat}(s_j) +  \nu_{18}{\rm Long}(s_j)) \times i^{2},
\end{align}
for $i \in \{1,...,h\}$; $\nu_{1},...,\nu_{18}$ are real numbers to be estimated.  
We work in relatively small areas and use distances in latitude and
longitude here and for the rest of this work.

\subsubsection{Conditional mean structure of $(\YObs|\Ynwp)$}\label{sec:condMean}
In \cite{Royle99a}, several configurations of the transition matrix $\Lambda$ are proposed depending on its use. 
For instance, a transition matrix from atmospheric pressure to wind speed is derived from geostrophic equations in \cite{Royle99b}.  
The observations exhibit daily and half-daily periodicity (with various intensities depending on the month of the year) and spatial patterns; see Figure \ref{fig:mean_hour}. 
However, the relation between the two datasets does not exhibit significant time dependence that requires a time-varying dependence. 
We use spatial and temporal neighbors to explain the observed wind speed. 
The land use LU is included in the transition matrix, because it defines different behaviors in the NWP model data. 
We choose the following transition between the two datasets: 
\begin{align}
\nonumber \mu_{Obs|NWP}(t,s) &= {\rm E}(\YObs(t,s)|\Ynwp) = \mu(t,s) + (\Lambda \Ynwp)(t,s)\,,~\textrm{with } \\  
\nonumber \mu(t,s) &= \Big(\beta_7 + \beta_8 \cos\Big(\frac{2\pi t}{24}\Big) + \beta_9 \sin\Big(\frac{2\pi t}{24}\Big) \\ 
\nonumber &  \qquad + \beta_{10} \cos\Big(\frac{2\pi t}{12}\Big) + \beta_{11} \sin\Big(\frac{2\pi t}{12}\Big) \Big)\\
\nonumber & \qquad \Big(1 + \alpha_{2}\rm{Lat}(s) + \alpha_{3}\rm{Long}(s)\Big), \\
\nonumber (\Lambda \Ynwp)(t,s) &=  \sum_{i=1}^{h} \rho^{(\rm{LU}(s))}(|t-t_{i}|)   \sum_{k=1}^{3} \Phi_{k}(\Delta \rm{Lat},\Delta \rm{Long})(s,s_{k})\Ynwp(t_{i},s_{k})\,, \\
\nonumber &  \qquad ~ t_{1} \leq t \leq t_{h}\,,
\end{align}
where 
\begin{description}
\item[-] $\rho^{(.)}(.)$ are temporal weights, parameterized according to $\rho^{(l)}(\Delta t)=\rho_{0}^{(l)}\exp(-\rho_{1}^{(l)}|\Delta t|)+\rho_{2}^{(l)}$, for the time difference $\Delta t$ in $\{0,..,h-1\}$;  
the integer $l \in \{1,..,n\}$ is the land use value of the closest grid point of $s$; 
 $\rho^{(l)}(0) = 1$ for identifiability purposes; 
\item[-] $\Phi_{\cdot}(\Delta \rm{Lat},\Delta \rm{Long}) = \phi_{0,.} + \phi_{1,.} \Delta \rm{Lat} + \phi_{2,.}\Delta \rm{Long} $, with $\Delta \rm{Lat}(s_{i},s_{j}) = |\rm{Lat}(s_{i})-\rm{Lat}(s_{j})|$ and  $\Delta \rm{Long}(s_{i},s_{j}) = |\rm{Long}(s_{i})-\rm{Long}(s_{j})|$; 
\item[-] $s_{1}$, $s_{2}$, $s_{3}$ are nearest spatial neighbor grid points of $s$ selected according to the radial distance, but other distances are possible. Moreover, for simplicity we consider here nearest neighbors, but other choices of predictors can be made, such as upwind stations;  and
\item[-] all the parameters $\beta_{7},...,\beta_{11}, \alpha_{2}, \alpha_{3}, (\rho_{0}^{(l)})_{l=1..n}, (\rho_{1}^{(l)})_{l=1..n}, (\rho_{2}^{(l)})_{l=1..n}, (\phi_{0,k})_{k=1..3} ,\\ (\phi_{1,k})_{k=1..3}, (\phi_{2,k})_{k=1..3} $ are real numbers to be estimated. 
\end{description}

\subsubsection{Conditional covariance structure of $(\YObs|\Ynwp)$} \label{sec:condVar}
Analysis of the empirical conditional covariance suggests the use of the parametric shape proposed in \eqref{eq:EqCov}, with a different set of parameters. 

\subsection{Estimation of the parameters}\label{sec:estimPar}

Maximum likelihood is chosen for estimating the parameters. 
The likelihood of the model for the observed dataset
$(\yobs(t_1,...,t_{k_{K}}; \allowbreak s_1,...,s_{J_{0}}),\\ \ynwp(t_1,...,t_{k_{K}};
\allowbreak s_1,...,s_{J_{0}}))$ is written as 
\begin{align*}
& \mathcal{L}(\theta ; \yobs(t_1,...,t_{k_{K}};s_1,...,s_{J_{0}}),\ynwp(t_1,...,t_{k_{K}};s_1,...,s_{J_{0}})) \\
&\qquad = p_{\theta}(\yobs(t_1,...,t_{k_{K}};s_1,...,s_{J_{0}}),\ynwp(t_1,...,t_{k_{K}};s_1,...,s_{J_{0}})) \\
&\qquad =  p_{\theta}(\ynwp(t_1,...,t_{k_{K}};s_1,...,s_{J_{0}})) \\ 
&\qquad \qquad p_{\theta}(\yobs(t_1,...,t_{k_{K}};s_1,...,s_{J_{0}})| \ynwp(t_1,...,t_{k_{K}};s_1,...,s_{J_{0}})).
\end{align*}
This is the particular instantiation of
\eqref{eq:abstract:likelihood}.

Each day, the WRF model is run independently from the previous day.  
Here we have assumed short temporal
  error correlations (see Fig. \ref{fig:map_corr}). Furthermore,
  the forecasts are restarted from reanalyses, and at least in the
  linear case the innovations are independent from observations \cite[\S 6.3]{Shumway10}. 
Therefore, we consider statistical independence between each day, which leads to the following product: 
\begin{align*}
 p_{\theta}(\ynwp(t_1,...,t_{k_{K}};s_1,...,s_{J_{0}}))  & =  \prod_{i=1}^{K}p_{\theta}(\ynwp(t_{k_{i}};\s),...,\ynwp(t_{k_{i}+23};\s)) \\
 & =   \prod_{i=1}^{K}p_{\theta}(\ynwp(b_{i};\s)), 
 \end{align*}
where $\s=\{s_1,...,s_{J_{0}}\}$ and $\{b_1,...,b_K\}=\{t_1,...,t_{24},t_{25},...,t_{k_{K}}\}$ with $b_i = \{t_{k_{i}},...,t_{k_{i}+23}\}$. 
For each $i \in \{1,..., K \}$ the associated log-likelihood is written as
\begin{align*}
\log( p_{\theta}(\ynwp(b_{i};\s)) ) = & -\frac{1}{2} \Big(  \log((2\pi)^{(24*J_{0})}) +\log(\det(\Sigma_{NWP})) \\ 
& + (\ynwp(b_{i};\s) - \mu_{NWP})^{T}\Sigma_{NWP}^{-1} (\ynwp(b_{i};\s) - \mu_{NWP}) \Big),
\end{align*}
where $\mu_{NWP}$ and $\Sigma_{NWP}$ are the parametric mean and covariance, respectively expressed in \eqref{eq:EqmuNWP} and \eqref{eq:EqCov}. 
The complete log-likelihood associated with the marginal distribution of $\Ynwp$ is then expressed as
\begin{align*}
& \log( p_{\theta}(\ynwp(t_1,...,t_{k_{K}};s_1,...,s_{J_{0}}))) \\ 
& =  -\frac{1}{2} \sum_{i=1}^{K} \Big( \log((2\pi)^{(24*J_{0})}) +\log(\det(\Sigma_{NWP})) \\ 
& \quad + (\ynwp(b_{i};\s) - \mu_{NWP})^{T}\Sigma_{NWP}^{-1} (\ynwp(b_{i};\s) - \mu_{NWP})\Big). 
\end{align*}
Similarly the conditional distribution is written as 
\begin{align*}
& \log( p_{\theta}(\yobs(t_1,...,t_{k_{K}};s_1,...,s_{J_0})| \ynwp(t_1,...,t_{k_{K}};s_1,...,s_{J_0}))) \\ 
& =  -\frac{1}{2} \sum_{i=1}^{K} \Big( \log((2\pi)^{(24*J_{0})}) +\log(\det(\Sigma_{Obs|NWP})) \\
& + (\yobs(b_{i};\s) - \mu - \Lambda \ynwp(b_{i};\s))^{T}\Sigma_{Obs|NWP}^{-1} (\yobs(b_{i};\s) - \mu - \Lambda \ynwp(b_{i};\s))\Big). 
\end{align*}
In practice, a preliminary least-squares estimation of the parameters is realized between the empirical and parametric first- and second-order structures of $\YObs$ and $\Ynwp$. 
These estimates are used as initial conditions of the maximum likelihood procedure.

\subsection{Kriging\label{sec:kriging}}
Predictions of $\YObs$ from $\Ynwp$ are obtained from the kriging equations \cite{Stein12}, with the mean and covariance defined by \eqref{fulldistri}.   
For $t_0$ in $b_{K+1}=\{t_{k_{K+1}},..., t_{k_{K+1}+h-1} \}$ and $s_0$ in $\{\s,\rm{S}_{0}\}=\{1,...,J_{0},J_{0}+1, ..., J\}$ defined in \eqref{EqSetup}, we have 
\begin{align}
\label{distribKrig}
(\YObs(t_0;s_0)|\Ynwp(b_{K+1};1,...,J_{0},J_{0}+1, ..., J)) \sim  \mathcal{N}\left(\hat \mu_{Obs} (t_0;s_0),\hat \Sigma_{Obs}(t_0;s_0) \right)
\end{align}
with 
\begin{subequations}
\begin{align}
\label{MuCovKrig}
& \hat \mu_{Obs} (t_0;s_0)  = (\mu+\Lambda \mu_{NWP})(t_0;s_0) + \\
\nonumber & \qquad \qquad c_{0}^{T} \Sigma_{NWP}^{-1}((b_{K+1};\s,\rm{S}_{0});(b_{K+1};\s,\rm{S}_{0})) ((\Ynwp-\mu_{NWP})(b_{K+1};\s,\rm{S}_{0}))\,,
\\
\hat \Sigma_{Obs}(t_0;s_0) & = \Sigma_{Obs}((t_0;s_0);(t_0;s_0)) +c_{0}^{T} \Sigma_{NWP}^{-1}((b_{K+1};\s,\rm{S}_{0});(b_{K+1};\s,\rm{S}_{0}))  c_0\,, \\
\nonumber c_0 & = \Sigma_{Obs,NWP}((t_0;s_0);(b_{K+1};\s,\rm{S}_{0})). 
\end{align}
\end{subequations}

The distribution \eqref{distribKrig} is used to generate the scenarios
of prediction of wind speed in Section \ref{SecRes}. This is in fact the
predictive distribution presented in \eqref{eq:pred:distr}.


\section{Wind data} \label{SecData}

In order to improve forecasts from the considered numerical model, two sources of data are combined: ground measurements and WRF model outputs.  
The measurement data are recorded across an irregular network, and at each observational station we pick the closest gridded point of NWP outputs. 
As a result, the two datasets have the same number of spatial locations; however, the proposed model is not restricted to this spatial layout and can handle datasets with different numbers of stations. 
In the following, the time series of the two datasets are filtered in time by a moving average process over a window of $1$ hour to remove small-scale effects and focus on a larger temporal scale; they are picked every hour.  
We focus on a region around Lake Michigan in the United States; however, the framework proposed here is not specific to that region. 

\subsection{Direct observations}
\begin{figure}[t]
  \centering
  \begin{tabular}{cc}
    \includegraphics[width=0.5\textwidth]{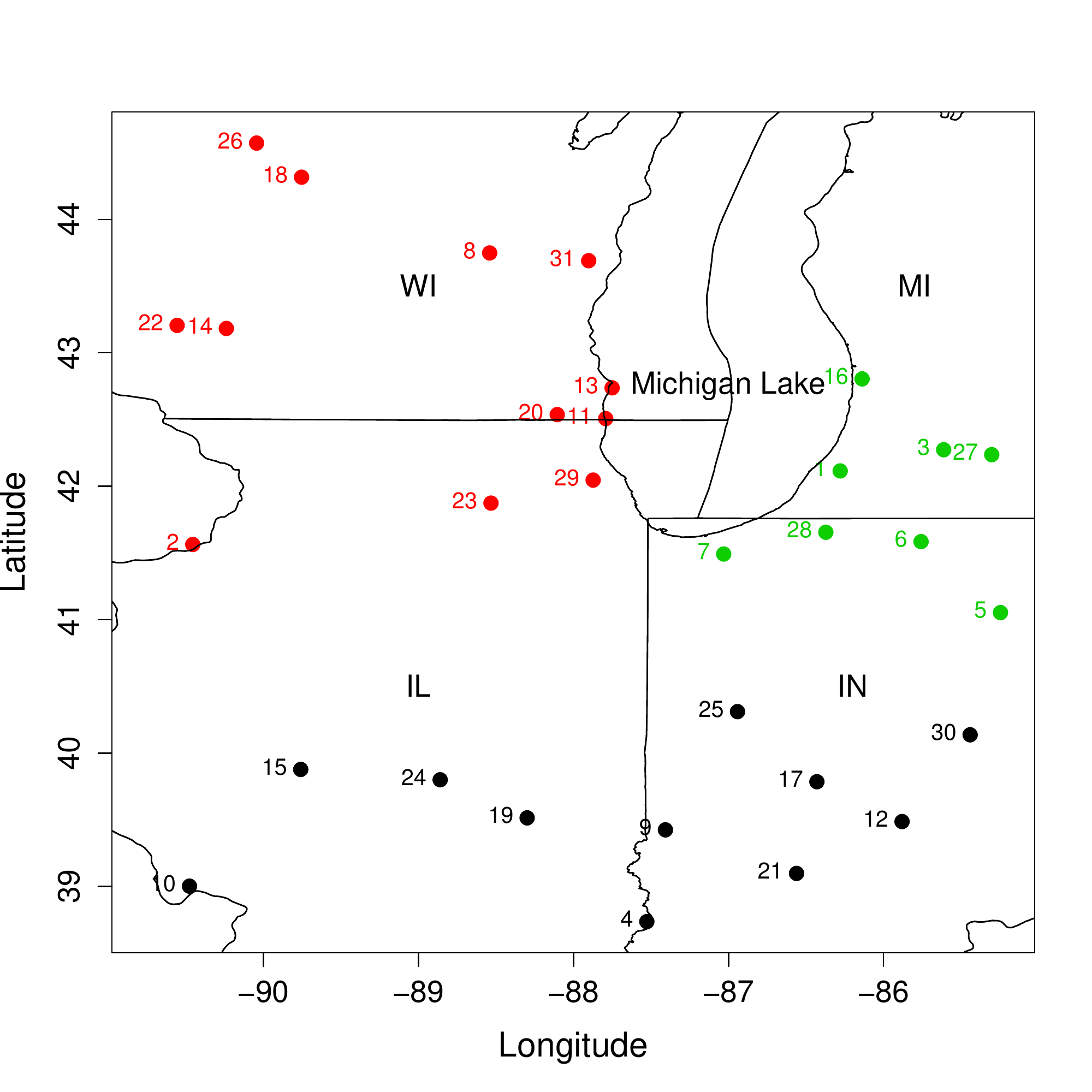}
    \end{tabular}
\caption{Map of the considered area (Midwest; visible are Lake
  Michigan, Michigan, Illinois, Indiana, and Wisconsin states): clusters are depicted with different colors, respectively in black are the 11 stations of sub-region $C_{1}$, in red the 12 stations of $C_{2}$ and in green the 8 stations of $C_{3}$}
  \label{fig:map_data}
\end{figure}

Observational data are extracted from the Automated Surface Observing System (ASOS) network, available at \url{ftp://ftp.ncdc.noaa.gov/pub/data/asos-onemin}. 
The network of collecting stations covers the U.S. territory.  
The studied data are 1-minute data selected from Wisconsin, Illinois, Indiana, and Michigan; see Figure  \ref{fig:map_data}.  
The measured wind speed is discretized in integer knots (one knot is about 0.5 m/s). 
We do not apply any additional treatment to account for this discretization because the data are filtered over a window of 1 hour; see \cite{Sloughter10} for a discussion of the discretization of the wind speed.  
The orography of this region is simple and flat; however, the presence of Lake Michigan has a strong impact on the wind conditions.  
Several months are investigated and reveal different behaviors; in particular, periodicities differ from winter to summer months.
In the following, for homogeneity purposes the dataset of 31 stations is subdivided into three spatial clusters of respectively 11, 12, and 8 stations, depicted in Figure \ref{fig:map_data}.  
A spatial clustering is performed on wind speed in order to distinguish among different average regional weather conditions. 
This is a proxy for different NWP forecast behaviors.  
These three clusters are treated independently hereafter. 

%
\subsection{Numerical weather prediction data}
%
State-of-the-art NWP forecasts are generated by using WRF v3.6
\cite{Skamarock08}, which is a state-of-the-art numerical
weather prediction system designed to serve both operational
forecasting and atmospheric research needs. WRF has a comprehensive
description of the atmospheric physics that includes cloud
parameterization, land-surface models, atmosphere-ocean coupling, and
broad radiation models. The terrain resolution can go up to 30 seconds
of a degree (less than $1~\textnormal{km}^2$). The NWP forecasts are
initialized by using the North American Regional Reanalysis fields dataset that covers the North American continent (160W-20W;
10N-80N) with a resolution of 10 minutes of a degree, 29 pressure
levels (1000-100 hPa, excluding the surface), every 3 hours from the year 
1979 until the present. Simulations are started every day during January
and August 2012 and cover the continental United States on a grid of 25x25 km
with a time resolution of 10 minutes.

\section{Results}\label{SecRes}

In this section, we first analyze the estimated parameters and then explore qualitatively and quantitatively the ability of the model to provide accurate forecasts.  
Two months of the year (January and August) are considered and are studied independently in order to investigate the model performance under different conditions.  
Moreover, the model is compared with two embedded models: one model with only temporal dependencies but without spatial interactions and one model without temporal or spatial dependencies.  
For each month, the model is trained on contiguous two-thirds of the month and predicted on the remaining third.  
The training periods are rolled over the three possible permutations
of one-third to fill in the entire month.

\subsection{Analysis of the estimated parameters}\label{subsec:ParamAnalysis}
\begin{figure}
\centering 
\includegraphics[scale=.3]{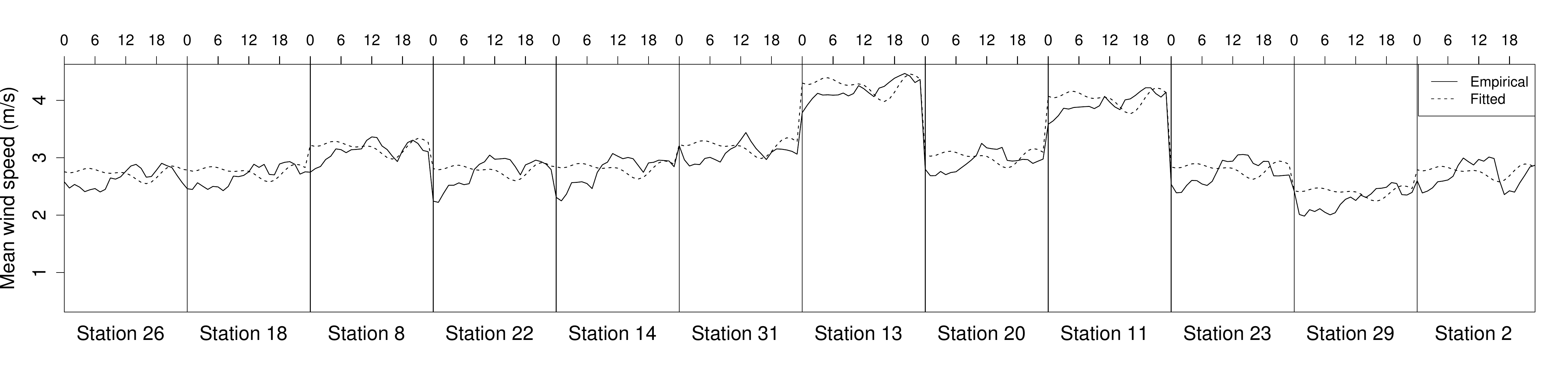}
\includegraphics[scale=.3]{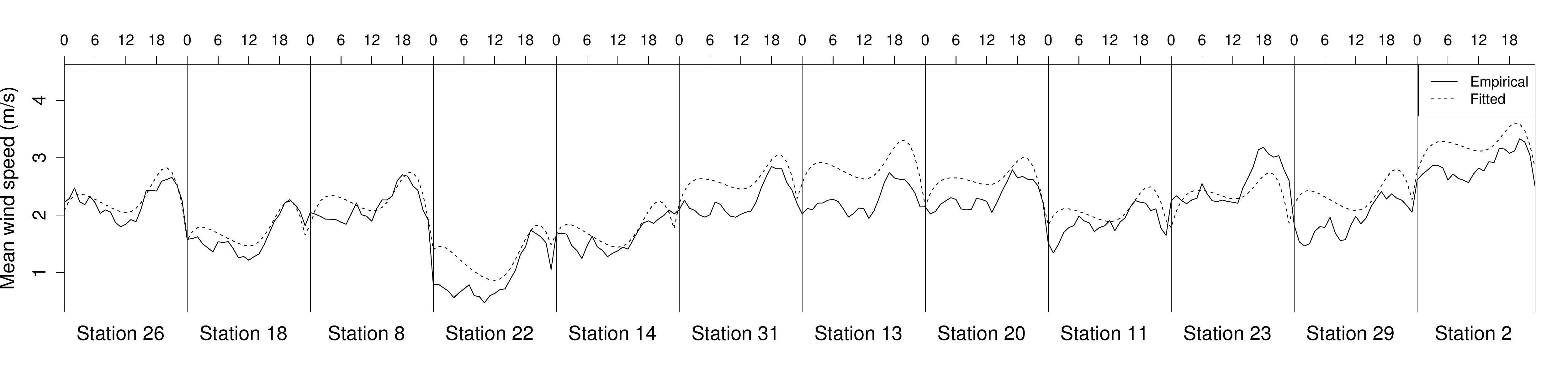}
\caption{Empirical and fitted parametric mean of wind speed at each hour of a day and at each station in the subregion C2 in January. Vertical lines separate each station. Within each of these windows, each hour of the day is considered. Top panel: mean of $\Ynwp$; bottom panel: mean of $\YObs$. }
\label{fig:mean_hour}
\end{figure}

\begin{figure}
\centering
\includegraphics[scale=.35]{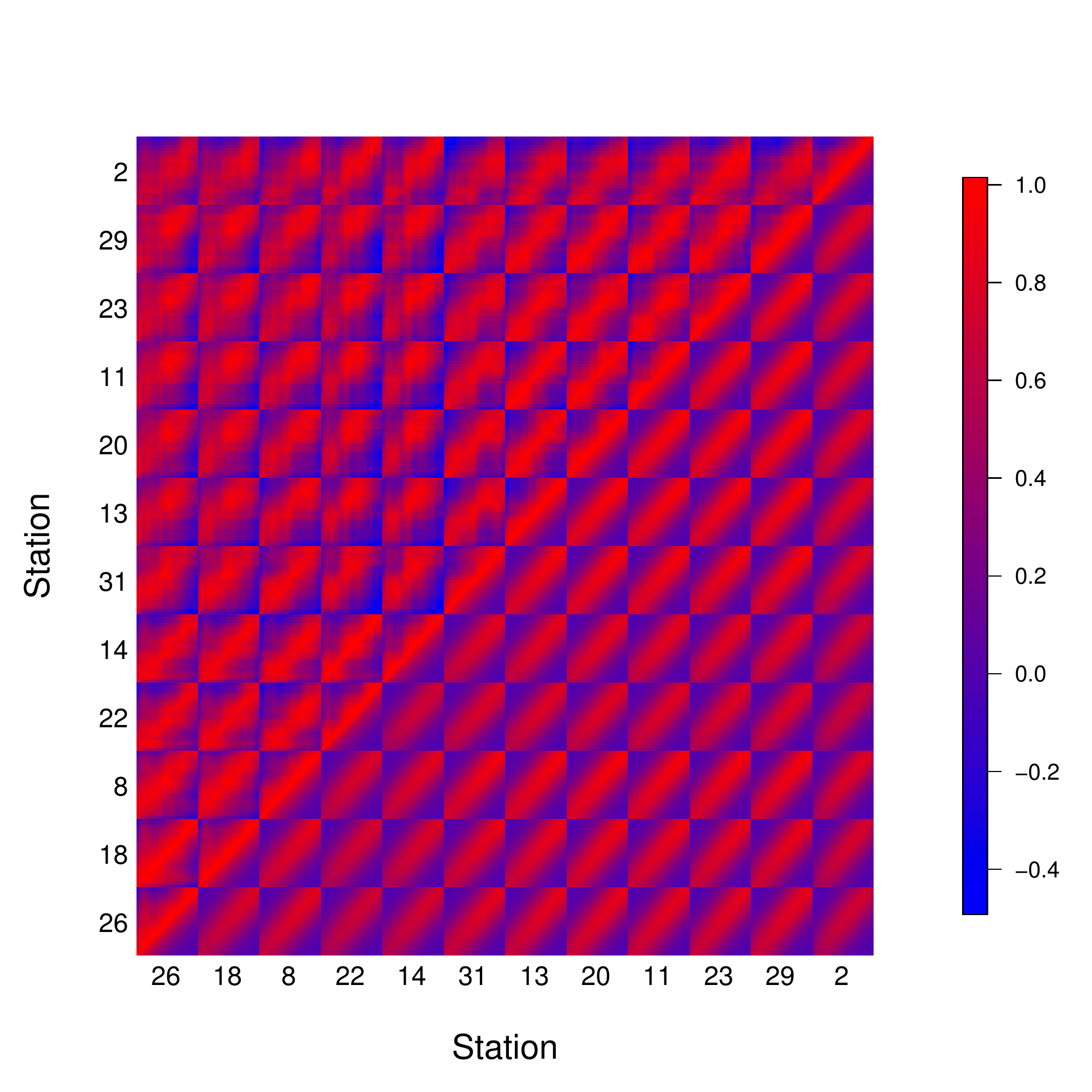}
\includegraphics[scale=.35]{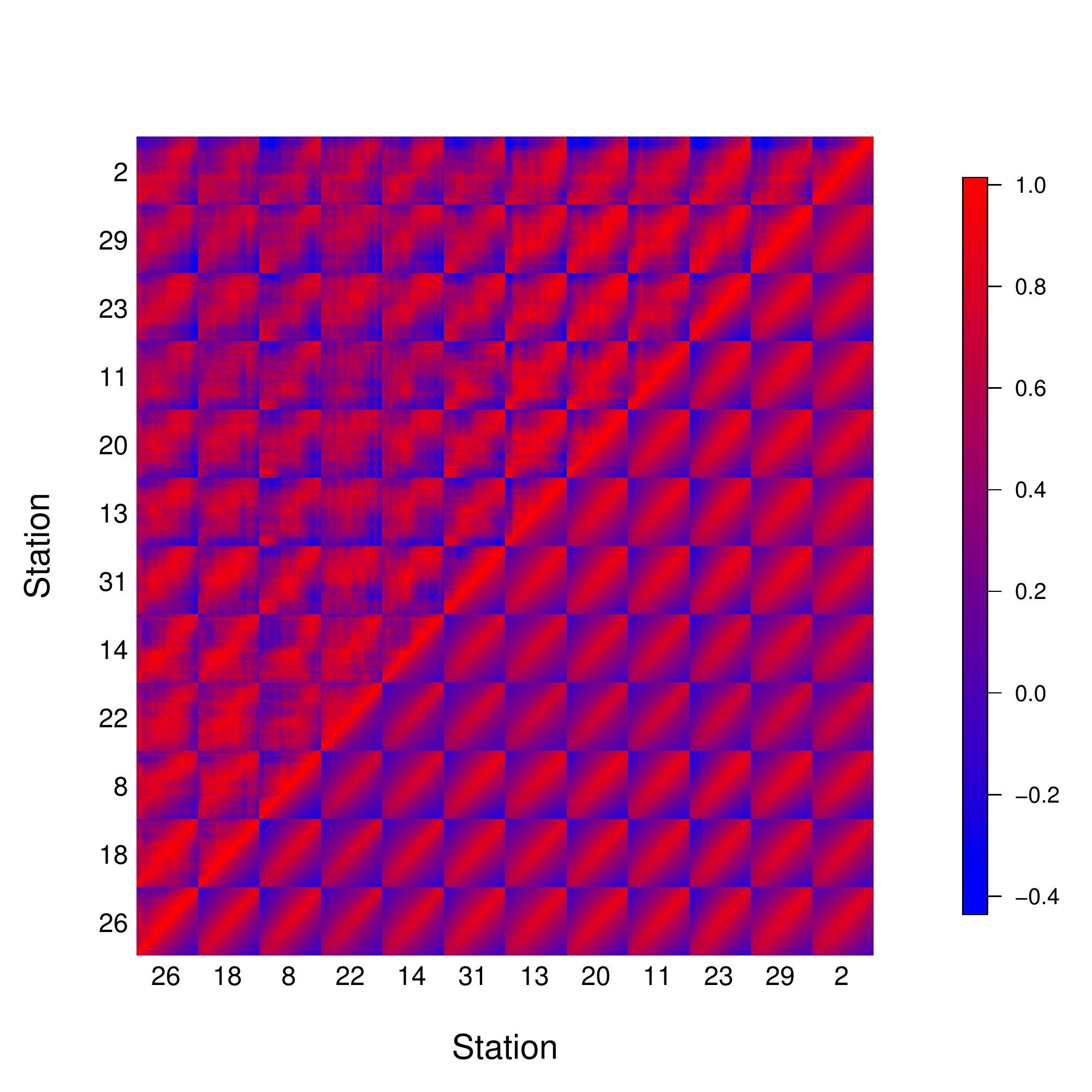}
\caption{Empirical and fitted parametric space-time correlation estimated in January 2012 in the subregion C2, left: NWP space-time correlation, right: measurement correlation. The upper left part of the matrix corresponds to the empirical correlation; the right bottom blocks display the fitted parametric correlation. Diagonal block correspond to temporal correlations at each station, and off-diagonal blocks correspond to temporal cross-correlation between stations. }
\label{fig:map_corr}
\end{figure}

In this section, we investigate the maximum likelihood estimation of the mean and covariance of the process.    
First, the empirical mean and covariance are compared with the fitted parametric ones proposed in Section \ref{SecModel}. 
The mean of the process $(\YObs,\Ynwp)$ is depicted in Figure \ref{fig:mean_hour}; for each station, the mean at each hour of the day is plotted. 
The structure of the estimated mean of the two processes is accurately reproduced in terms of temporal and spatial patterns. 
In Figure \ref{fig:map_corr}, the empirical and fitted space-time correlations are plotted. 
A great part of the structure is captured by the proposed parametric shapes; however, the global shapes tend to be smoothed by the parametric models. 
The nonseparability between space and time that is visible on the empirical off-diagonal blocks is not entirely captured by the parametric model on the top panels. 
Analysis of the matrices $\PropYbase_{s}$ that are involved in the
covariance model \eqref{eq:EqCov} reveals different configurations given the subregion and the period of the year. 
These can be expected because these operators can be interpreted as a linear projector of a process that is common to all the stations. 
Average air flows differ according to the season and the location; the dependence from a common process that would contain this information is likely to differ in space and in time across the year. 

The matrix $\Lambda$, which appears in both the mean and covariance components, is important because it links the NWP forecasts to the objective predictive quantities. 
The analysis of $\Lambda$ reveals that the intensity of temporal
dependence varies with the land use; however, the temporal persistence
is curtailed to a few hours across the different land use.

\begin{figure}
\centering 
\includegraphics[scale=.37]{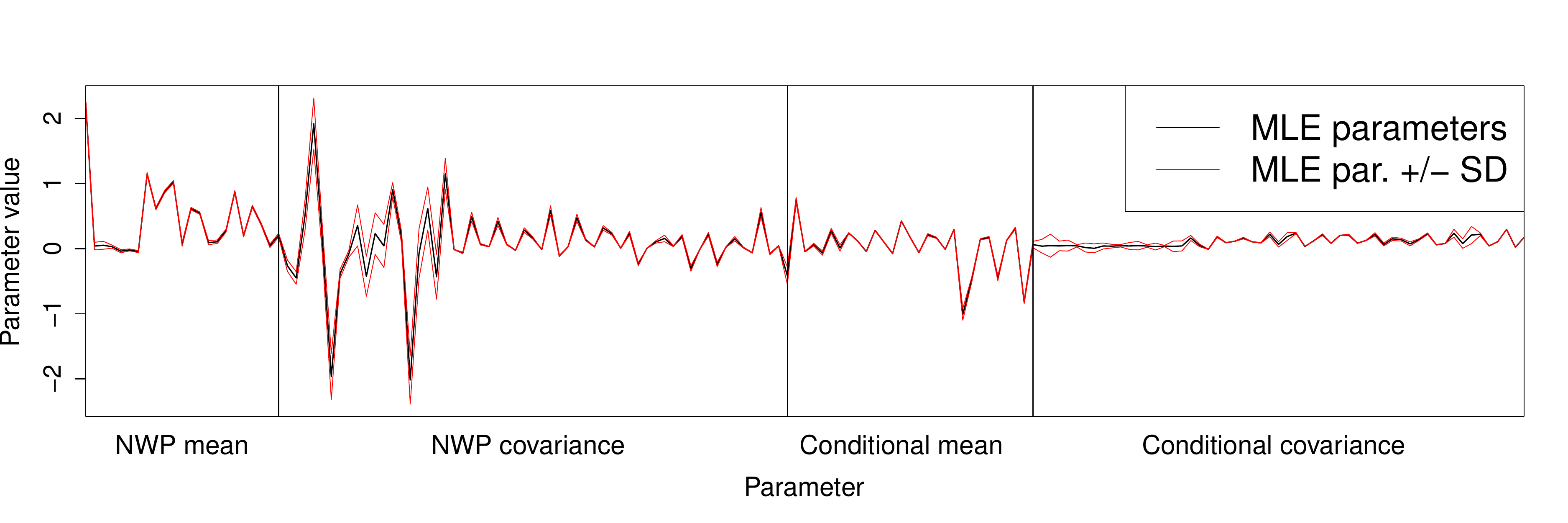}
\caption{Maximum likelihood estimation of the parameters of the model and associated standard deviation of estimation. Vertical lines separate the different sets of parameters. From left to right are the parameters of $\mu_{NWP}$, $\Sigma_{NWP}$, $\mu_{Obs|NWP}$, and $\Sigma_{Obs|NWP}$. Parameters are plotted following this order: 
$\mu_{NWP}$: $(\beta_{k})_{k=0,...,6},(\alpha_{0}^{(l)})_{l=1,...,n}, (\alpha_{1}(j))_{j=1,...,J_{0}}$, 
$\Sigma_{NWP}$: $\nu_{1},...,\nu_{18},\sigma_{0}, \lambda_{0},\gamma_{0},(\sigma_{j},\lambda_{j},\gamma_{j})_{j=1..J_{0}}$, 
$\mu_{Obs|NWP}$: $\beta_{7},...,\beta_{11}, \alpha_{2}, \alpha_{3}, (\rho_{0}^{(l)})_{l=1..n}, (\rho_{1}^{(l)})_{l=1..n}, (\rho_{2}^{(l)})_{l=1..n}, (\phi_{0,k})_{k=1..3}, (\phi_{1,k})_{k=1..3},\\ (\phi_{2,k})_{k=1..3}$, 
and $\Sigma_{Obs|NWP}$: same indexation as $\Sigma_{NWP}$. }
\label{fig:parameteruncert}
\end{figure}

In the second step, the uncertainty associated with the estimation of the parameters is accounted for. 
In Figure \ref{fig:parameteruncert}, the maximum likelihood estimation of the parameters and the associated standard deviation of estimation, given by the inverse of the Hessian of the log-likelihood calculated at the maximum, are plotted.  
Note that the uncertainty is relatively narrow as we assumed previously. 
The greatest estimation variance is presented by several parameters $\nu_{i}$ that appear in the matrices $\PropYbase$ in Section \ref{sec:covNWP}.  
In the parameters of $\mu_{Obs|NWP}$, parameters with a high estimation variance are the ones associated with $\mu$ defined in Section \ref{sec:condMean}. 
In these cases, a lack of data in the estimation of these specific parameters may cause this high estimation variance.

\subsection{Assessment of the quality of the predictive model}\label{subsec:TimePred}

In this part, samples (or scenarios) are generated from the predictive distribution defined by Equation \eqref{distribKrig}.  
The mean of these samples can be used as a pointwise prediction, but the objective here is to embed the uncertainty associated with the prediction by working with samples from the predictive distribution.  The presented predictive scenarios are back-transformed by using the inverse Box-Cox transformation.    

We compare the model proposed in Section \ref{SecModel} with two reductions of it: a model where only temporal dependence is accounted for and spatial interactions are ignored and another reduction where only the bias in mean in corrected, and neither temporal nor spatial dependencies are accounted for.  

\subsubsection{Qualitative exploration of the predictions}

\begin{figure}
\centering
\includegraphics[scale=.23]{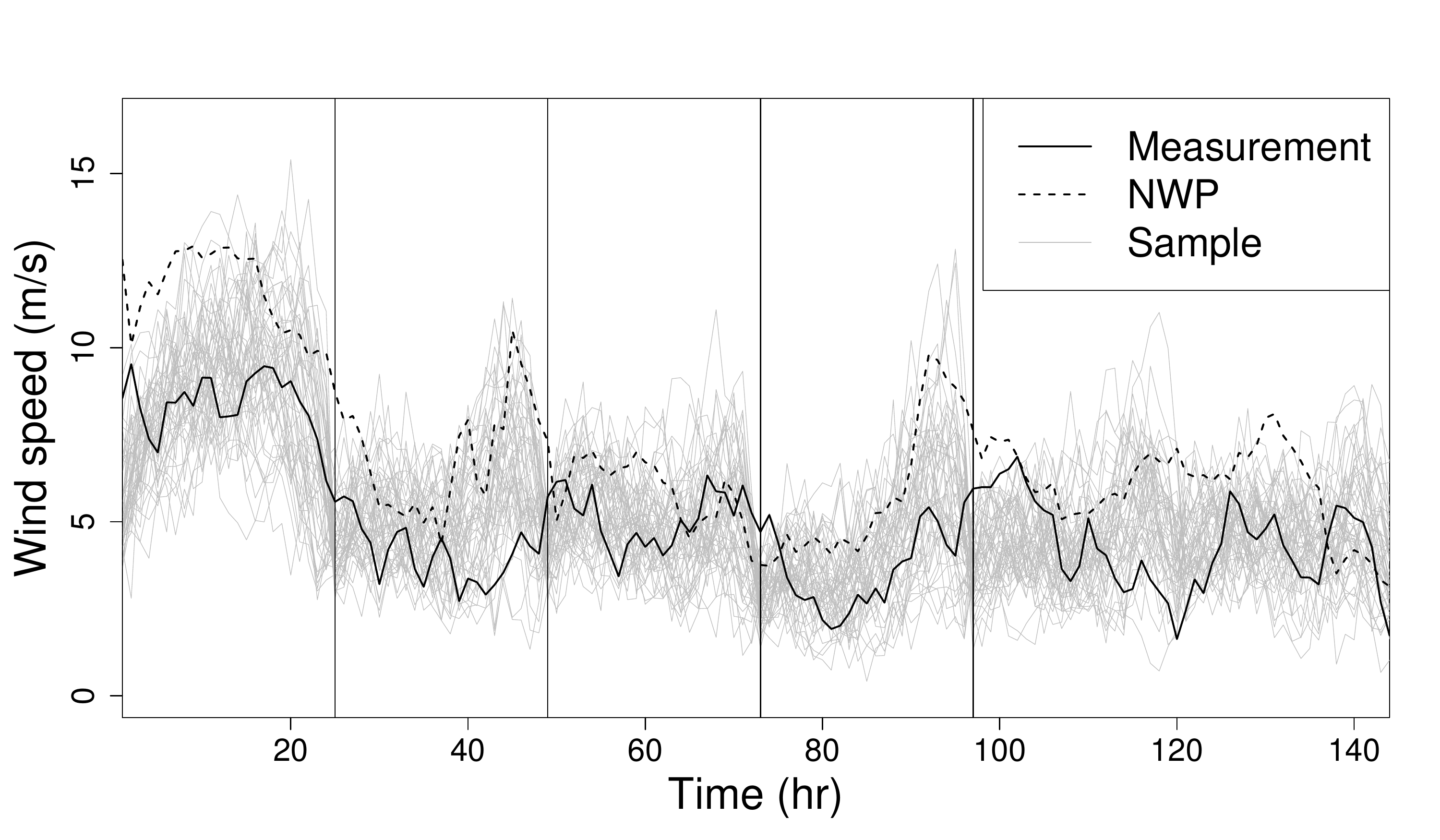}
\hglue.00001cm
\includegraphics[scale=.23]{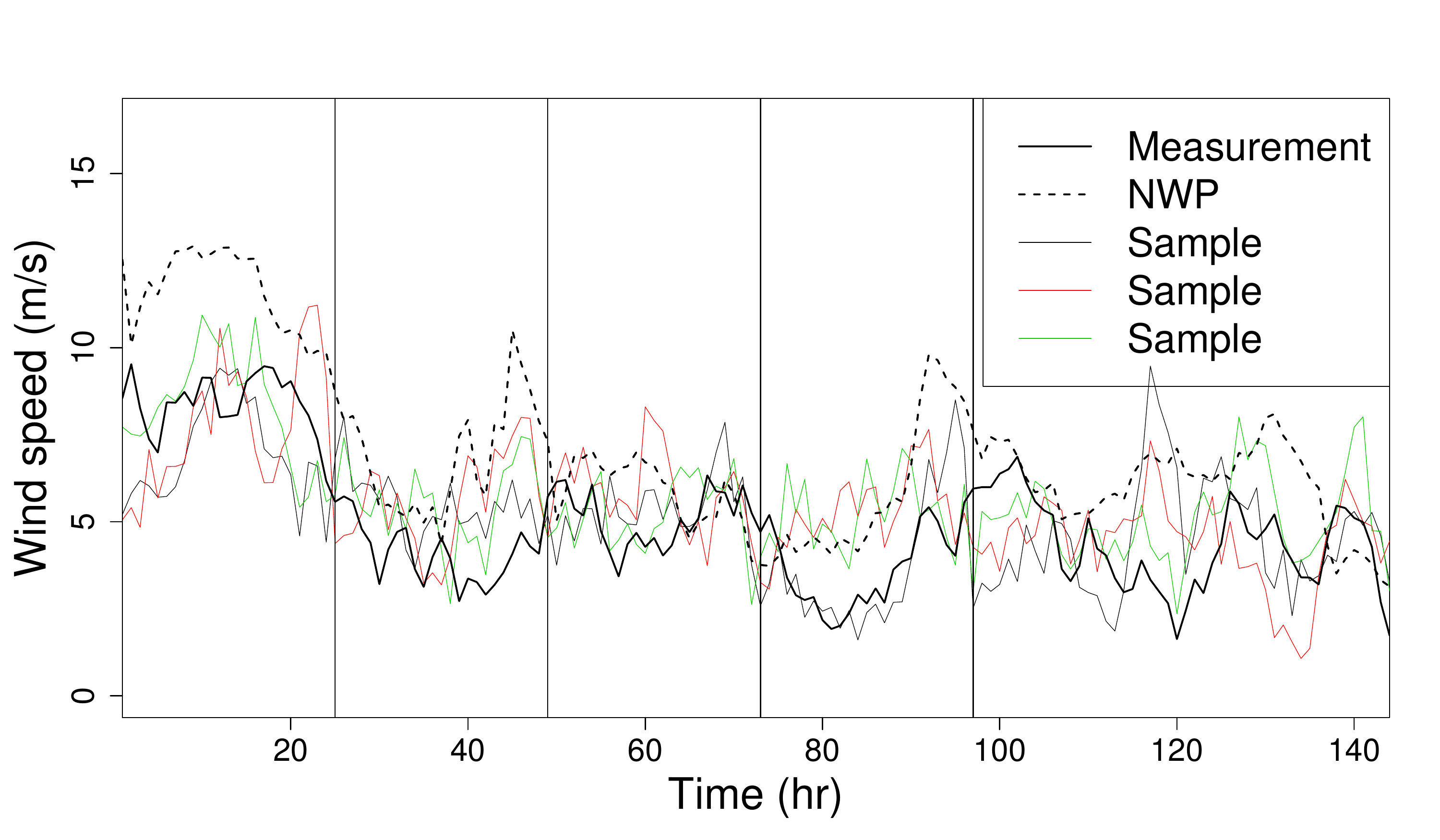}
\vglue.01cm
\includegraphics[scale=.23]{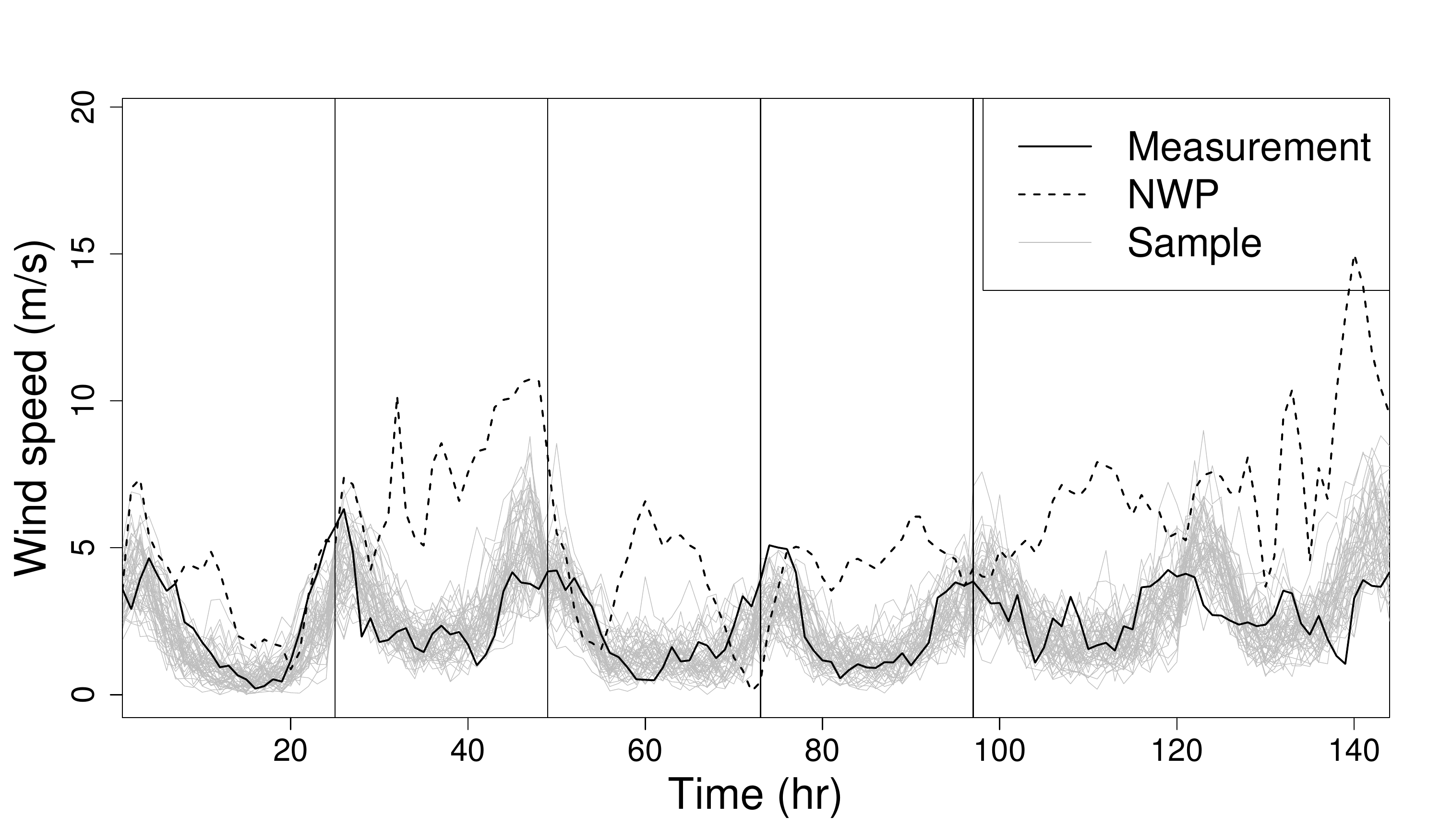}
\hglue.00001cm
\includegraphics[scale=.23]{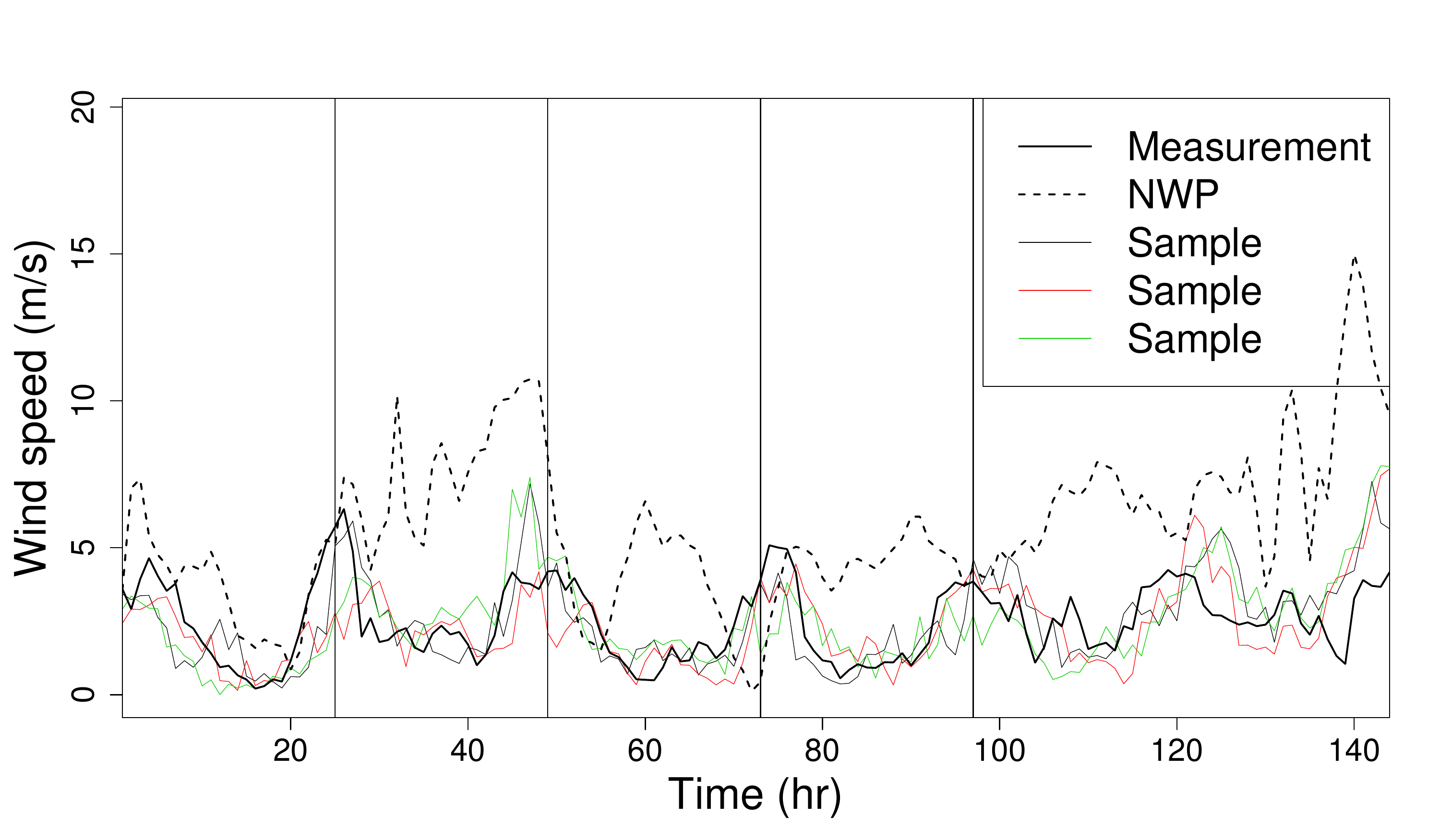}
\caption{Time series of wind speed at the station with the median
RMSE in subregion C2. January 2012 (top) and August 2012 (bottom) for six days that are separated by vertical lines. Left panels: 50 predictive samples are plotted; right panel: 3 samples are plotted.}
\label{fig:timeseries}
\end{figure}

\begin{figure}
\centering
\includegraphics[scale=.28]{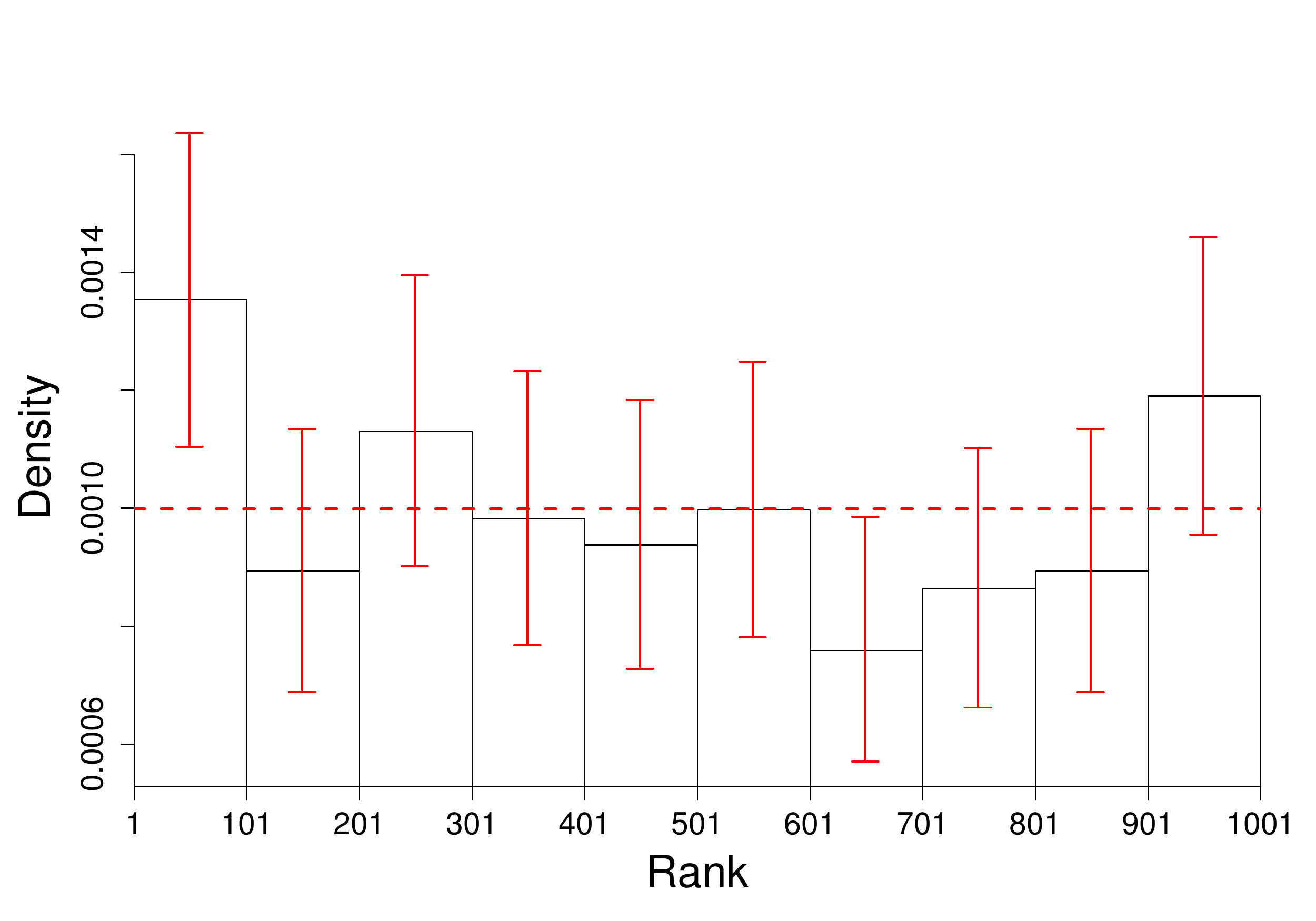}
\includegraphics[scale=.28]{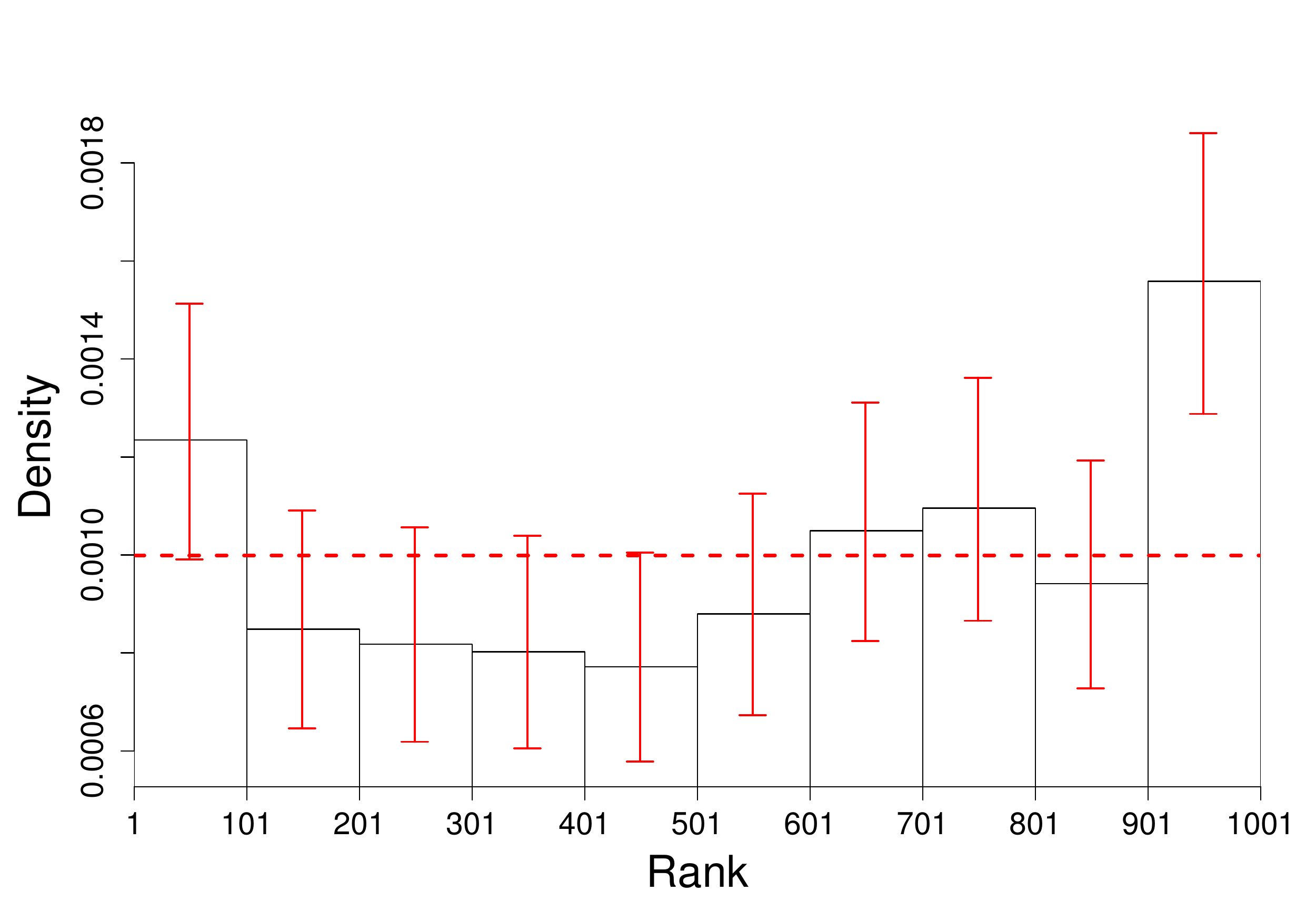}
\caption{Univariate rank histograms at the station with the median RMSE in subregion C2, left: January 2012, right: August 2012. The red error bars are $95\%$-confidence intervals (confidence intervals are computed with a binomial distribution); the horizontal red line represents the density of a uniform distribution.} 
\label{fig:rankhisto}
\end{figure}

\begin{figure}
\centering 
\includegraphics[scale=.2]{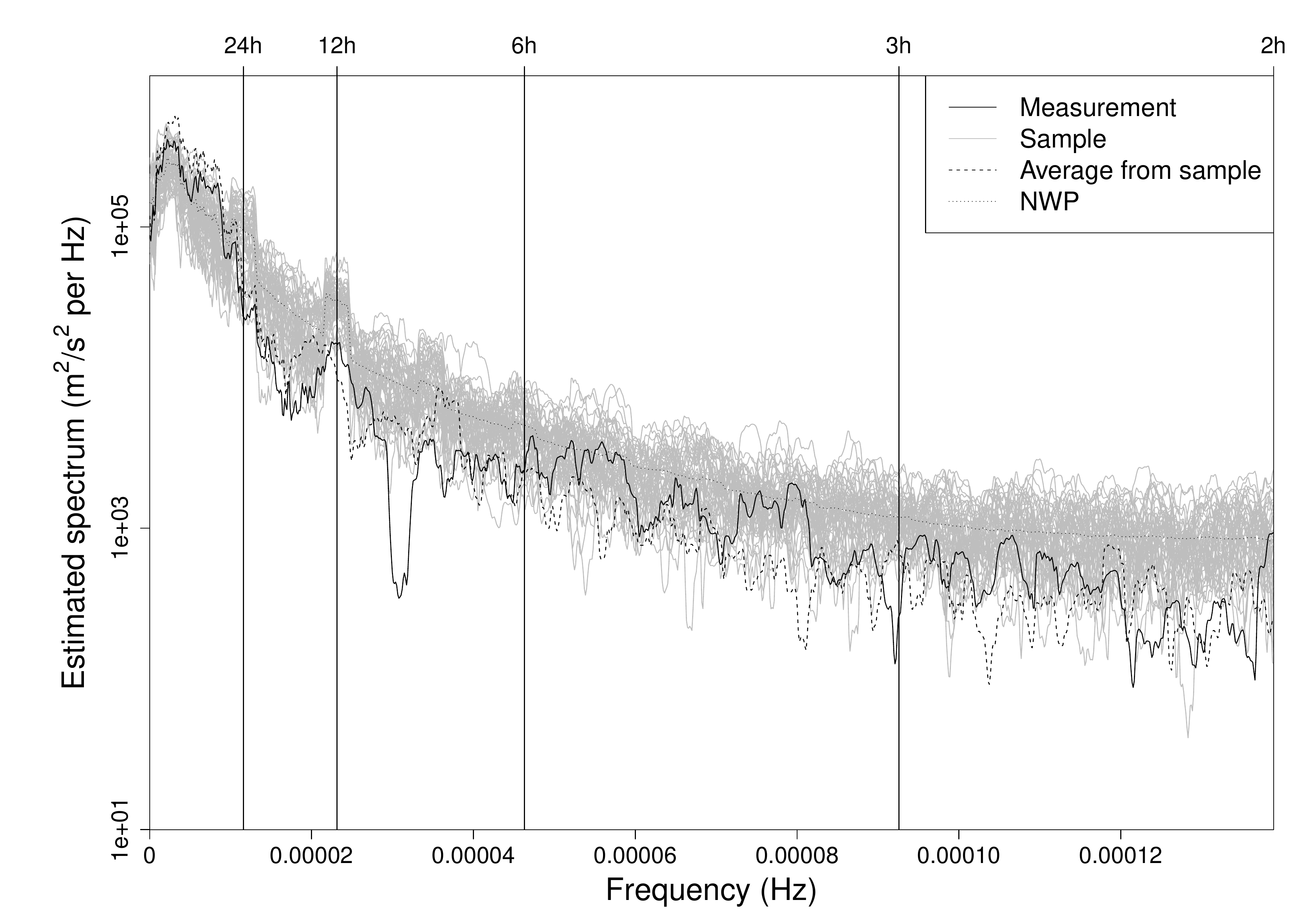}
\includegraphics[scale=.2]{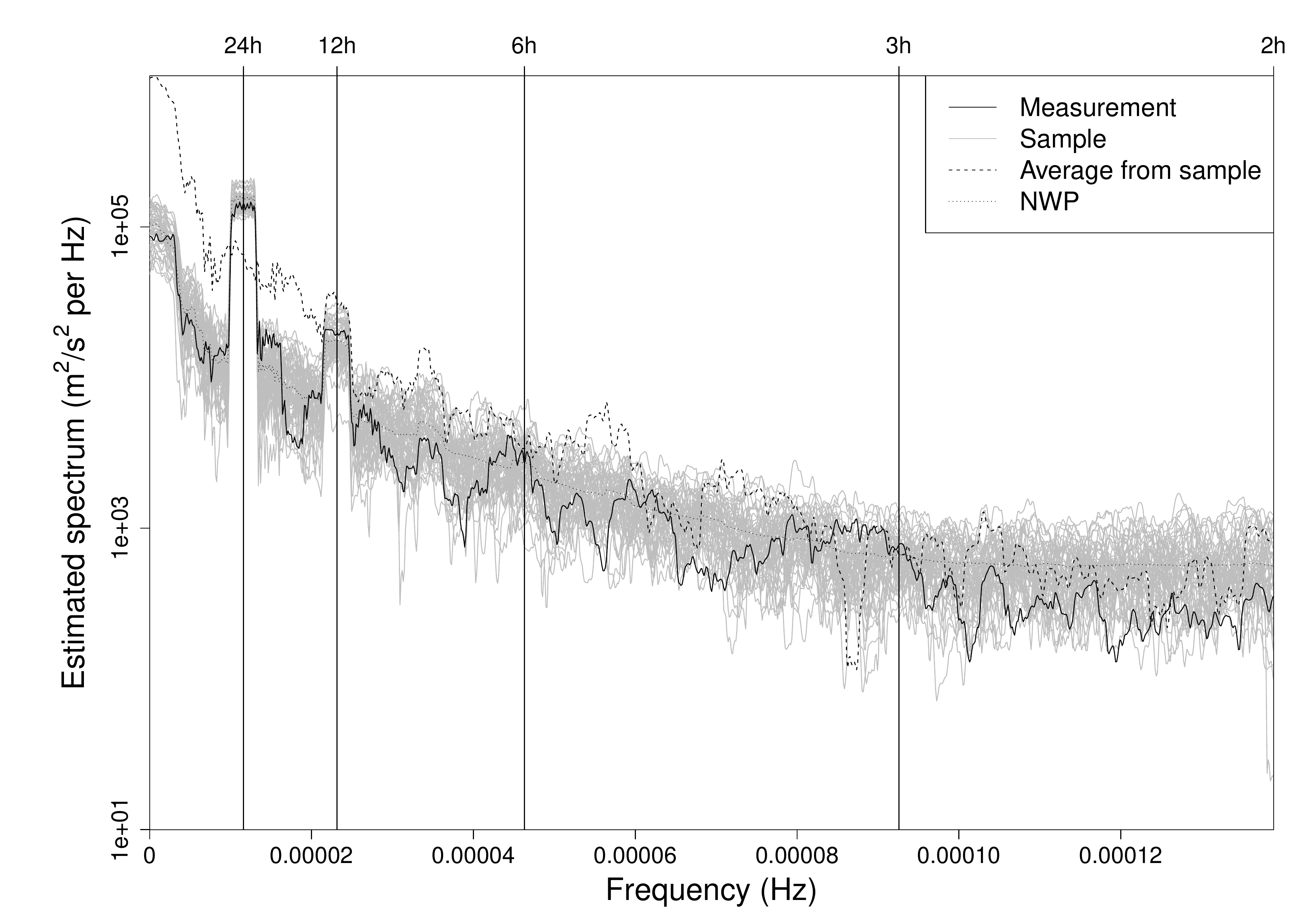}
\caption{Estimated spectrum in January (left) and August (right) for the station with the median RMSE in subregion C2. }
\label{fig:spectrum}
\end{figure}

\begin{figure}
\centering
\includegraphics[scale=.23]{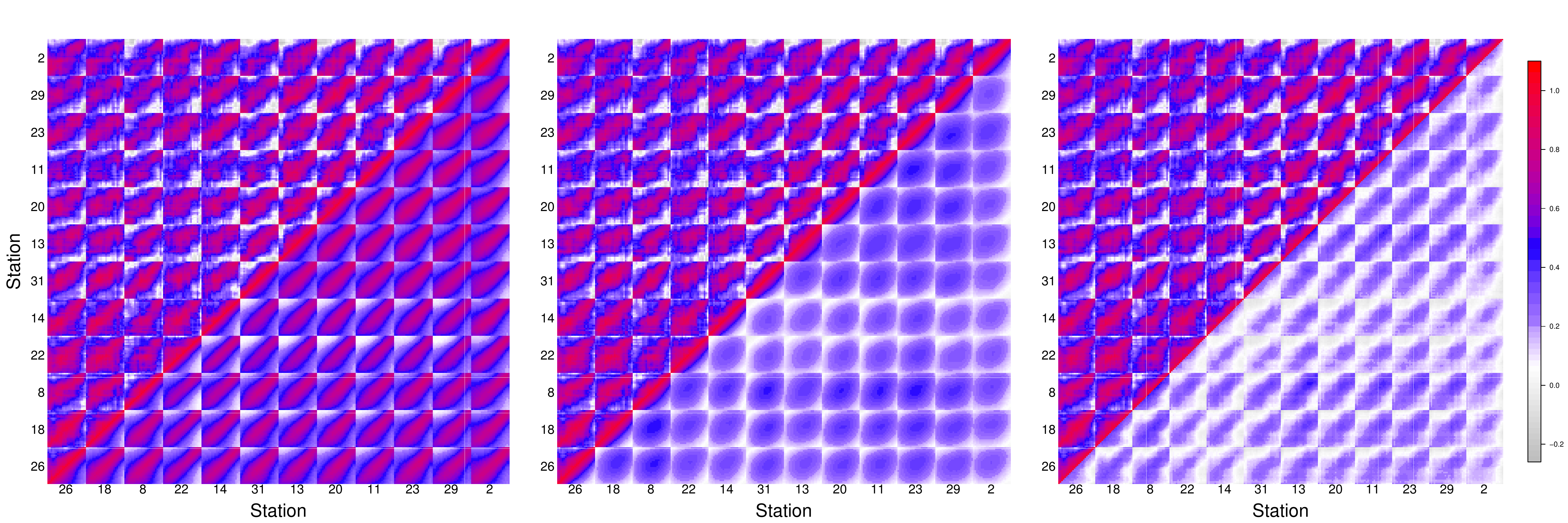}
\caption{Space-time correlation of wind speed in January in sub-region $C_{2}$. The upper left part of the matrix corresponds to the empirical correlation, the right bottom blocks display the fitted correlation. Diagonal block correspond to temporal correlations at each station, and off-diagonal blocks correspond to temporal cross-correlation between stations. From left to right: full space-time model, temporal model, bias-correction model. }
\label{fig:corr_spacetime}
\end{figure}

We first investigate the time series of the predictions by a visual assessment in Figure \ref{fig:timeseries}. 
We next explore how the forecast scenarios are representative in terms of calibration through rank histograms (Figure \ref{fig:rankhisto}) and in terms of temporal and spatio-temporal structures through spectral and correlation investigations (Figures \ref{fig:spectrum} and \ref{fig:corr_spacetime}).   

Diagnostic tools and scores to evaluate vector-valued prediction have been proposed more recently for single and ensemble forecasts \cite{Smith04,Gneiting08,Pinson12,Thorarinsdottir14,Scheuerer15b}.  
Some of these criteria evaluate simultaneously univariate predictive skills and multivariate dependencies. 
Criteria for ensemble forecasts can be used for predictive scenarios where each sample of the predictive distribution is treated as a member of the ensemble \cite{Pinson12}.  
Here we focus on univariate criteria for  the calibration separately from the temporal and spatio-temporal dependencies; however, diagnostic tools such as multivariate rank histograms  \cite{Gneiting08,Thorarinsdottir14} can be used to assess univariate predictive skills and multivariate dynamics. 
Multivariate scores are discussed and compared in Subsection \ref{sec:metrics}.

\paragraph{Visual assessment of time series}
We investigate observed time series and generated predictive scenarios for part of the months of January and August; see Figure \ref{fig:timeseries}.  
Measured wind speed, which is to be predicted, is plotted as a reference in order to evaluate the accuracy of the prediction. 
NWP wind forecasts are also plotted because they are predictors and a target to be improved with respect to the measurements. 
For both months, the global trend of the measured time series is well captured by the predictive mean and by the scenarios. 
The predictive samples cover the measurements that are to be predicted (see left panels); and the predictive mean realizes, most of the time, an improvement with respect to the NWP forecasts.  
Moreover, each sample has a temporal dynamics consistent with the observed temporal behavior (right panels).  
The scenarios take negative values; however, such values  happen only $0.9\%$ of the time in January and $0.08\%$ in August. 
The improvement of the proposed prediction is more visible in August (bottom panels), likely because of the periodic components that are stronger in this period of the year and that are well captured by the model; see also Figure \ref{fig:spectrum}.  
Furthermore, the spread of the scenarios is more important in January
than in August, likely because the wind speed has
more variability in winter, as illustrated in the observed variances
in Table \ref{tab:MetricsForecast}, which may make it less predictable.  
We note that the scenarios are not spreading at the end of each
prediction window, as observed in the literature.  
The reason is that the NWP predictors are available over the entire prediction
window and such spread increase is not obvious in the
model--measurement discrepancy. 
%

\paragraph{Calibration assessment}
Rank histograms are commonly used to assess the calibration of predictive ensembles; they can be seen as analogous to ensembles of the probability integral transform (PIT) that evaluates the calibration of single forecasts \cite{Anderson96,Hamill01}. 
In Figure \ref{fig:rankhisto}, univariate rank histograms are plotted for one station, and each predictive scenario is treated as an ensemble member in the rank histogram. 
No signs of trend or over- and under- dispersion are seen in these histograms.
The two panels indicate a well-calibrated ensemble of samples; the horizontal line representing the uniform distribution is covered most of the time by the confidence intervals associated with the proportions of the histogram. 
The predictions from January tend to show better calibration than the August ones do.

\paragraph{Temporal spectral analysis}
The spectral content of the scenarios and of the observations is
estimated and depicted in Figure \ref{fig:spectrum}; the average
spectrum of the estimated spectrum on each sample is also plotted. We
note that a similar analysis was carried out in \cite{Bouallegue15}.
The estimated spectra of the scenarios cover most of the spectrum of the observations. 
The overall shape of the estimated spectrum and of the average
spectrum indicates a robust agreement, especially in August where small
frequencies are accurately captured. In this and other
spectral estimates, the spectral content at high frequency is
sometimes slightly overpredicted; we believe the reason is that the
time series has 
discontinuities at the boundaries between temporal
blocks due to overlapping prediction windows, which in turn are a
result of assimilating data and restarting the NWP forecast. Nevertheless, the features of the spectrum of the
measurements appear well captured by our model. 
Notice that the spectrum of NWP data in August provides a poor
description of the observations; however, the model is able to correct this. 
Therefore, our model appears to be a suitable
and realistic wind scenario generator. 
%

\paragraph{Space-time correlation structure}
In Figure \ref{fig:corr_spacetime}, empirical space-time correlations of measurements are compared with those of scenarios from the model proposed in Section \ref{SecModel} and with those of scenarios from the simplified models. 
The space-time model enables of the space-time correlation structure to be captured, whereas the embedded models capture significantly less information than does the full model. 
These two embedded models miss most of the spatial cross-correlation between stations. 
From this figure we see the importance of space-time information in the structure of the wind speed prediction.

\subsubsection{Quantitative assessment of the quality of the predictions}\label{sec:metrics}

As our second step, we assess quantitatively the overall improvement of the model in comparison with the WRF model outputs.  
We study general metrics because we would like to preserve a general
application scope. 
See \cite{Pinson13} for reflections on links between improvement of these general metrics and user-specific metrics and also for general challenges associated with forecast verification. 

\paragraph{Univariate predictive skills}

In Table \ref{tab:MetricsForecast}, the root mean square error (RMSE) is computed for the predictive mean of the proposed distribution and for the NWP forecasts. 
We consider also the energy score (ES), which represents a generalization of the continuous ranked probability score (CRPS) for ensemble predictions (see \cite{Gneiting08,Pinson12}). 
This metric is an omnibus metric that enables comparison of ensemble forecasts and scenarios with pointwise prediction; it is computed on predictive samples and on NWP forecasts. 
The energy score is a proper scoring rule, the lower the energy score, the better the proposed forecast. 
In subregion $C_2$, the model shows the greatest improvement in terms
of RMSE and energy score, likely because of the influence of Lake
Michigan on the results. 
Indeed, the NWP embeds this presence through the lake mask and land use, but this may be overestimated in comparison with the behaviors of the observations. 
The improvement in terms of  RMSE is more significant in August, likely because of the periodic components that are well captured by the model.  
The energy score clearly favors the proposed model in comparison with the WRF outputs. 
Most of the means and variances of the observations are well captured by the prediction made with the model.  
In addition, in Table \ref{tab:MetricsForecast}, the metrics are computed for the full space-time model and for its two simplifications (temporal model and bias-correction model). 
The proposed space-time model reveals better results than do the embedded models, as expected. 
Results are presented only for January in sub-region $C_{2}$; however, similar conclusions can be drawn from the other months and sub-regions. 
\newcommand{\coltbl}{0.75}
\begin{table}
\centering
\begin{tabular}{|c|c|c||c|c|c|}
\hline
Model & RMSE $(m/s)$ & ES $(m/s)$ & M/S & Mean($\YObs$)  $(m/s)$ & Var($\YObs$) $(m^{2}/s^{2})$ \\
\hline
 \rowcolor[gray]{\coltbl}
NWP (Jan. 2012, $C_1$) & 3.03 & 78 & M & 3.72 & 4.15 \\
\hline
Model (Jan. 2012, $C_1$) &  1.61 (47$\%$) & 30 & S & 4.32 & 4.63 \\
\hline
 \rowcolor[gray]{\coltbl}
NWP (Aug. 2012, $C_1$) & 1.62& 41 & M & 2.55 & 2.29 \\
\hline
Model (Aug. 2012, $C_1$)  & 1.07 (34$\%$) & 19 & S & 2.29 & 2.32 \\
\hline
\hline
 \rowcolor[gray]{\coltbl}
NWP (Jan. 2012, $C_2$)  & 2.61 & 67 & M & 4.44 & 5.31  \\
\hline
Model (Jan. 2012, $C_2$) & 1.67 (36$\%$) & 31 & S & 4.56 & 5.37  \\
\hline
 \rowcolor[gray]{\coltbl}
NWP (Aug. 2012, $C_2$)  & 4.03 & 102 & M & 2.5 & 2.3  \\
\hline
Model (Aug. 2012, $C_2$) &  1.05 (74$\%$)  & 18 & S & 2.4 & 2.42 \\
\hline
\hline
 \rowcolor[gray]{\coltbl}
NWP (Jan. 2012, $C_3$)  & 2.31 & 60 & M & 4.72 & 6.54  \\
\hline
Model (Jan. 2012, $C_3$) & 1.85 (20$\%$) & 34 & S & 4.86 & 5.4  \\
\hline
 \rowcolor[gray]{\coltbl}
NWP (Aug. 2012, $C_3$)  & 1.72 & 44 & M & 2.31 & 2.23  \\
\hline
Model (Aug. 2012, $C_3$) &  1.02 (40$\%$)  & 18 & S & 2.31 & 2.18 \\
\hline
\hline
 \rowcolor[gray]{\coltbl}
NWP (Jan. 2012, $C_2$)  & 2.61 & 67 & M & 4.44 & 5.31  \\
\hline
Model (Jan. 2012, $C_2$) & 1.67 (36$\%$) & 31 & S & 4.56 & 5.37  \\
\hline 
\hline
\rowcolor[gray]{0.9}
Model Temp. (Jan. 2012, $C_2$) & 1.88 (28$\%$) & 33 & S & 3.74 & 4.72  \\
\hline
\rowcolor[gray]{0.9}
Model Bias (Jan. 2012, $C_2$) & 1.95 (25.5$\%$) & 34 & S & 3.68 & 4.4  \\
\hline
\end{tabular}
\caption{Statistics and metrics for the station, representing the median RMSE in each cluster denoted as $C_i$, for  $i=1,2,3$. 
ES is the energy score and M/S is measurements or samples.  
They are evaluated on the concerned month for time prediction. 
Associated with the model RMSE is the percentage of improvement of the model with respect to the NWP data. 
The bottom block corresponds to metrics computed on the three proposed reductions of the model. }
\label{tab:MetricsForecast}
\end{table}
\begin{figure}
\centering
\includegraphics[scale=.33]{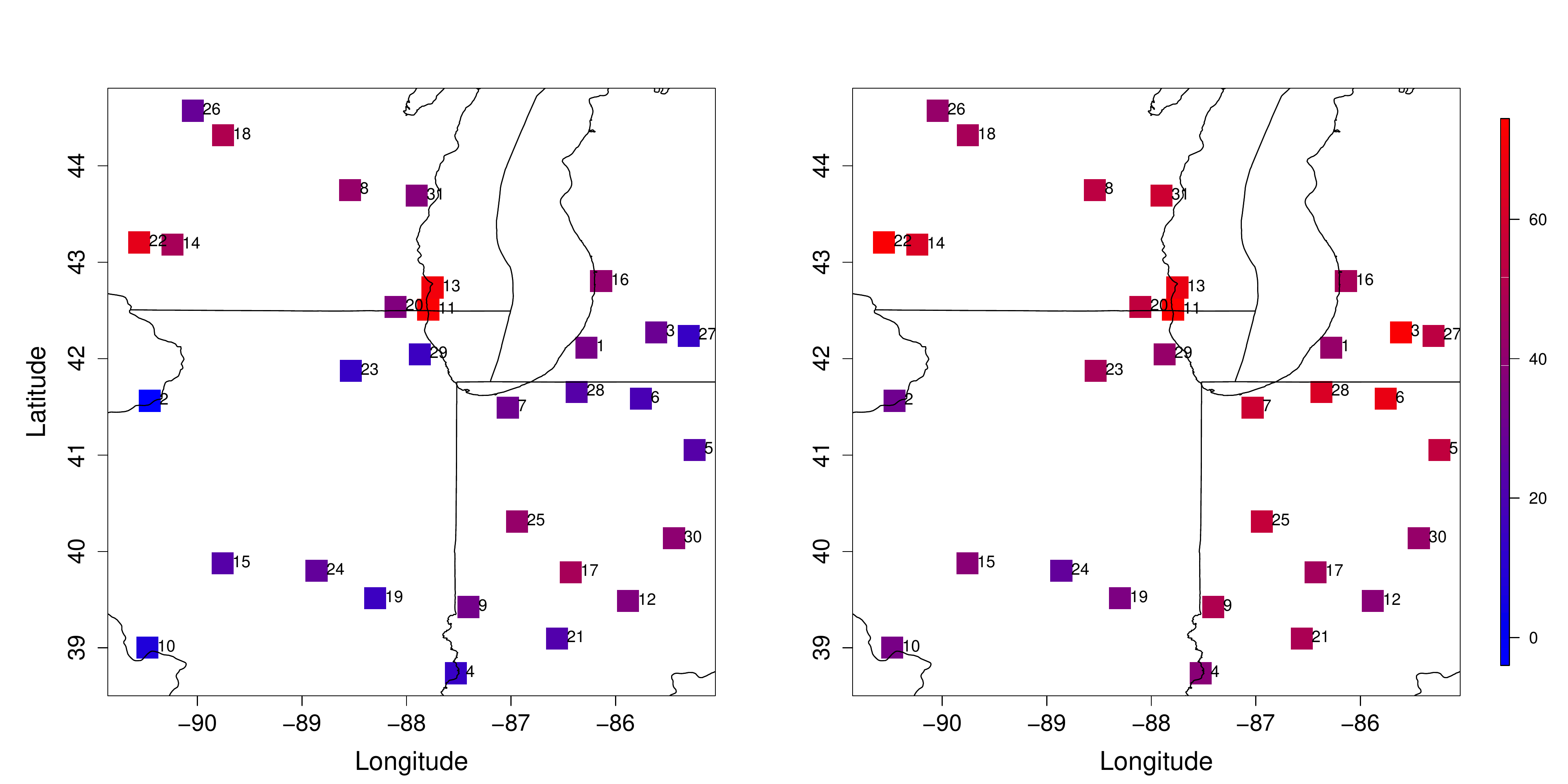}
\caption{Maps of percentage of improvement of RMSE of the proposed prediction with respect to RMSE of NWP prediction, left: January, right: August. }
\label{fig:map_improv}
\end{figure}
In Figure \ref{fig:map_improv}, maps of the percentage of improvement of RMSE are shown for each station and for both months. 
The improvement is greater in August and also around Lake Michigan in sub-region $C_2$.

\begin{figure}
\centering
\includegraphics[scale=.5]{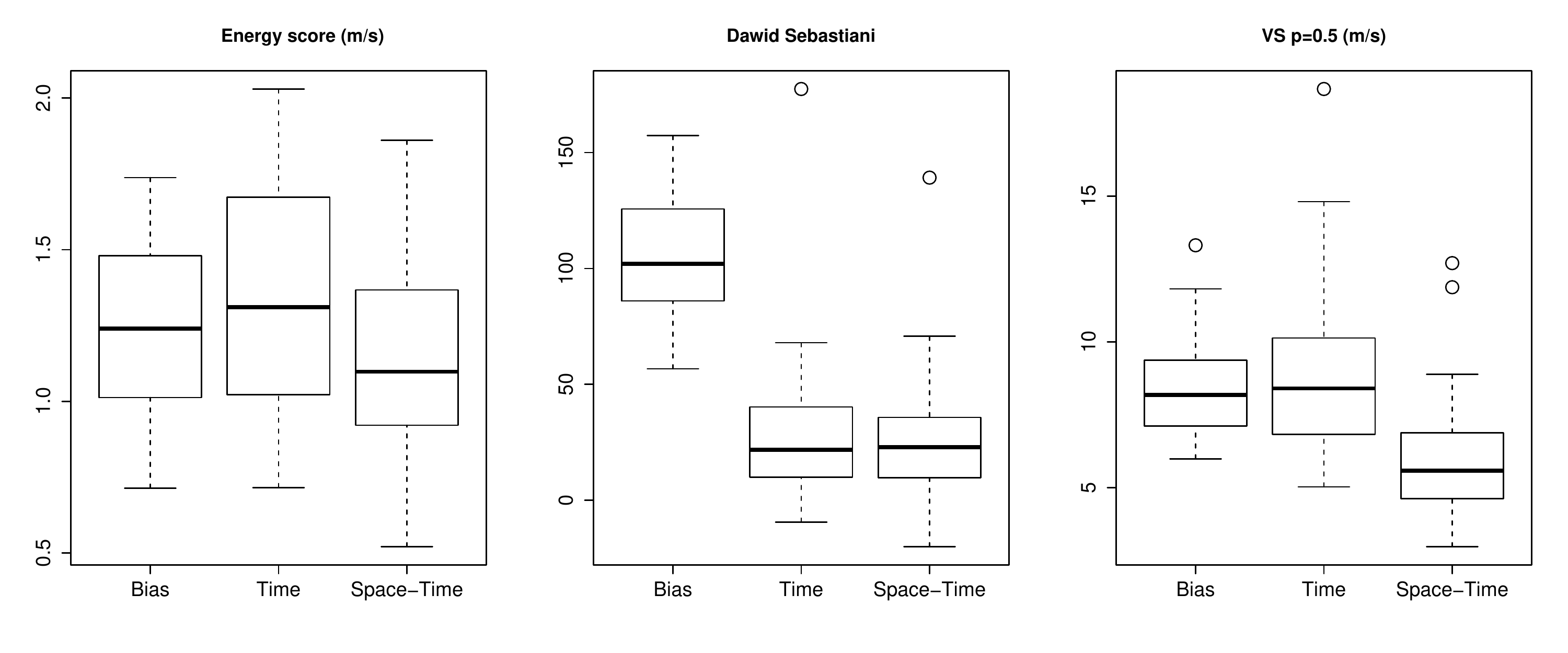}
\caption{Energy score, Dawid-Sebastiani score, and Variogram-based score computed for the station with median RMSE in sub-region $C_{2}$ and its closest neighbor for 24 hours, for January 2012. The variogram-based score is computed with equal weights and for an order $p=0.5$, which shows here the most discrimination between the models. Scores are computed for the proposed model and its two reductions: a temporal model and a bias-correction model. Each point of the boxplot corresponds to the score computed for one day of the month. }
\label{fig:MVscores}
\end{figure}

\paragraph{Multivariate predictive skills}
In Figure \ref{fig:MVscores}, we compute three multivariate proper scores for the proposed model and its two reductions, namely the energy score; the Dawid-Sebastiani score \cite{Dawid99}, which is equivalent to the log-score for a Gaussian predictive distribution, and the variogram-based score \cite{Scheuerer15b}, which measures the dissimilarity between variograms of the observation and of the forecasts. 
The distributions of the scores computed for every day of the month of January are displayed in the boxplots of Figure \ref{fig:MVscores}. 
The energy score may not be discriminative in the case of mis-specified correlations; see \citep{Pinson13b,Scheuerer15b} and Figure \ref{fig:MVscores}.  
However, it helps discriminate accurate intensity in forecasts; see Table \ref{tab:MetricsForecast}.  
Note that in Figure \ref{fig:MVscores} the values of the energy score differ from those in Table \ref{tab:MetricsForecast} because different windows are under consideration: a space-time window is used in Figure \ref{fig:MVscores} and a temporal one in Table \ref{tab:MetricsForecast}.  
The Dawid-Sebastiani score enables  the model that corrects only the bias to be strictly distinguished from the models that embed also the temporal and space-time structures. 
The variogram-based score enables the space-time model to be distinguished from the two others, but it does not discriminate the temporal model from the bias-correction one.    
These multivariate proper scores assess different properties of the predictions and exhibit different results.    
A sensible approach, therefore, is to combine several scores in order to select appropriate predictive multivariate  models and assess their skills.  

A potential limitation of our approach stems from the
assumption that an inherent stationarity exists across the
calibration and forecast windows. In our case all days of a third of
each month are modeled with the same model. We currently are exploring
introducing non-stationarity between each temporal block (day) in
order to account for such potential shortcomings.

\section{Conclusions} \label{SecConclu}

We have introduced a statistical space-time modeling framework for
predicting atmospheric wind speed based on deterministic numerical weather
predictions and historical measurements. 
We have used a Gaussian
multivariate space-time process that combines multiple sources of
past physical model outputs and measurements along with model
predictions to forecast wind speed at observation sites.  
We applied this strategy to surface wind-speed forecasts for a region
near the U.S. Great Lakes. 
The results show that the
prediction is improved in the mean-squared sense as well as in
probabilistic scores. 
Moreover, the samples are  shown to
produce realistic wind scenarios based on the sample spectrum.  
Using the proposed model, one can correct the first- and second-order
space-time structure of the numerical forecasts in order to match the structure
of the measurements.

\section*{Acknowledgements}
This material was based upon work supported by the U.S. Department of Energy, Office of Science through
contract no. DE-AC02-06CH11357. We thank Prof. Michael Stein for
comments on multiple versions of this manuscript. 
We thank Michael Scheuerer for providing us codes on multivariate scores and his helpful comments.

\bibliographystyle{apalike}
\bibliography{biblio_manuscrit.bib}

\begin{thebibliography}{}

\bibitem[Ailliot et~al., 2006]{Ailliot06drift}
Ailliot, P., Fr\'enod, E., and Monbet, V. (2006).
\newblock Long term object drift forecast in the ocean with tide and wind.
\newblock {\em Multiscale Modeling and Simulation}, 5(2):514--531.

\bibitem[Anderson, 1996]{Anderson96}
Anderson, J.~L. (1996).
\newblock A method for producing and evaluating probabilistic forecasts from
  ensemble model integrations.
\newblock {\em Journal of Climate}, 9(7):1518--1530.

\bibitem[Anitescu et~al., 2012]{Anitescu12}
Anitescu, M., Chen, J., and Wang, L. (2012).
\newblock A matrix-free approach for solving the parametric gaussian process
  maximum likelihood problem.
\newblock {\em SIAM Journal on Scientific Computing}, 34(1):A240--A262.

\bibitem[Apanasovich and Genton, 2010]{Apanasovich10}
Apanasovich, T.~V. and Genton, M.~G. (2010).
\newblock Cross-covariance functions for multivariate random fields based on
  latent dimensions.
\newblock {\em Biometrika}, 97(1):15--30.

\bibitem[Bao et~al., 2010]{Bao10}
Bao, L., Gneiting, T., Grimit, E.~P., Guttorp, P., and Raftery, A.~E. (2010).
\newblock Bias correction and {B}ayesian model averaging for ensemble forecasts
  of surface wind direction.
\newblock {\em Monthly Weather Review}, 138(5):1811--1821.

\bibitem[Baran, 2014]{Baran14a}
Baran, S. (2014).
\newblock Probabilistic wind speed forecasting using bayesian model averaging
  with truncated normal components.
\newblock {\em Computational Statistics \& Data Analysis}, 75:227--238.

\bibitem[Baran and Lerch, 2014]{Baran14b}
Baran, S. and Lerch, S. (2014).
\newblock Log-normal distribution based emos models for probabilistic wind
  speed forecasting.
\newblock {\em arXiv preprint arXiv:1407.3252}.

\bibitem[Baran and Lerch, 2015]{Baran15}
Baran, S. and Lerch, S. (2015).
\newblock Mixture emos model for calibrating ensemble forecasts of wind speed.
\newblock {\em arXiv preprint arXiv:1507.06517}.

\bibitem[Berliner, 2000]{Berliner00}
Berliner, M. (2000).
\newblock Hierarchical {B}ayesian modeling in the environmental sciences.
\newblock {\em AStA Advances in Statistical Analysis}, 2(84).

\bibitem[Berrocal et~al., 2012]{Berrocal12}
Berrocal, V.~J., Gelfand, A.~E., and Holland, D.~M. (2012).
\newblock Space-time data fusion under error in computer model output: an
  application to modeling air quality.
\newblock {\em Biometrics}, 68(3):837--848.

\bibitem[Bouallegue et~al., 2015]{Bouallegue15}
Bouallegue, Z.~B., Heppelmann, T., Theis, S.~E., and Pinson, P. (2015).
\newblock Generation of scenarios from calibrated ensemble forecasts with a
  dynamic ensemble copula coupling approach.
\newblock {\em arXiv preprint arXiv:1511.05877}.

\bibitem[Bourotte et~al., 2015]{Bourotte15}
Bourotte, M., Allard, D., and Porcu, E. (2015).
\newblock A flexible class of non-separable cross-covariance functions for
  multivariate space-time data.
\newblock {\em Preprint}.

\bibitem[Brisson et~al., 2003]{Brisson03}
Brisson, N., Gary, C., Justes, E., Roche, R., Mary, B., Ripoche, D., Zimmer,
  D., Sierra, J., Bertuzzi, P., and Burger, P. (2003).
\newblock An overview of the crop model stics.
\newblock {\em European Journal of Agronomy}, 18(3):309--332.

\bibitem[Brown et~al., 1984]{Brown84}
Brown, B.~G., Katz, R.~W., and Murphy, A.~H. (1984).
\newblock Time series models to simulate and forecast wind speed and wind
  power.
\newblock {\em Journal of Climate and Applied Meteorology}, 23:1184--1195.

\bibitem[Constantinescu and Anitescu, 2013]{Constantinescu13}
Constantinescu, E. and Anitescu, M. (2013).
\newblock Physics-based covariance models for {G}aussian processes with
  multiple outputs.
\newblock {\em International Journal for Uncertainty Quantification},
  3(1):47--71.

\bibitem[Constantinescu et~al., 2011]{Constantinescu11}
Constantinescu, E., Zavala, V., Rocklin, M., Lee, S., and Anitescu, M. (2011).
\newblock A computational framework for uncertainty quantification and
  stochastic optimization in unit commitment with wind power generation.
\newblock {\em IEEE Transactions on Power Systems}, 26(1):431--441.

\bibitem[Cowles et~al., 2002]{Cowles02}
Cowles, M.~K., Zimmerman, D.~L., Christ, A., and McGinnis, D.~L. (2002).
\newblock Combining snow water equivalent data from multiple sources to
  estimate spatio-temporal trends and compare measurement systems.
\newblock {\em Journal of Agricultural, Biological, and Environmental
  Statistics}, 7(4):536--557.

\bibitem[Cressie and Wikle, 2011]{Cressie11}
Cressie, N. and Wikle, C.~K. (2011).
\newblock {\em Statistics for spatio-temporal data}.
\newblock John Wiley \& Sons.

\bibitem[Dawid and Sebastiani, 1999]{Dawid99}
Dawid, A.~P. and Sebastiani, P. (1999).
\newblock Coherent dispersion criteria for optimal experimental design.
\newblock {\em Annals of Statistics}, pages 65--81.

\bibitem[Fanshawe and Diggle, 2012]{Fanshawe12}
Fanshawe, T.~R. and Diggle, P.~J. (2012).
\newblock Bivariate geostatistical modelling: a review and an application to
  spatial variation in radon concentrations.
\newblock {\em Environmental and Ecological Statistics}, 19(2):139--160.

\bibitem[Feldmann et~al., 2014]{Feldmann14}
Feldmann, K., Scheuerer, M., and Thorarinsdottir, T.~L. (2014).
\newblock Spatial postprocessing of ensemble forecasts for temperature using
  nonhomogeneous gaussian regression.
\newblock {\em arXiv preprint arXiv:1407.0058}.

\bibitem[Fuentes et~al., 2005]{Fuentes05}
Fuentes, M., Chen, L., Davis, J.~M., and Lackmann, G.~M. (2005).
\newblock Modeling and predicting complex space--time structures and patterns
  of coastal wind fields.
\newblock {\em Environmetrics}, 16(5):449--464.

\bibitem[Fuentes and Raftery, 2005]{Fuentes05b}
Fuentes, M. and Raftery, A.~E. (2005).
\newblock Model evaluation and spatial interpolation by {B}ayesian combination
  of observations with outputs from numerical models.
\newblock {\em Biometrics}, 61(1):36--45.

\bibitem[Gel et~al., 2004]{Gel04}
Gel, Y., Raftery, A.~E., and Gneiting, T. (2004).
\newblock Calibrated probabilistic mesoscale weather field forecasting: The
  geostatistical output perturbation method.
\newblock {\em Journal of the American Statistical Association},
  99(467):575--583.

\bibitem[Genton and Kleiber, 2014]{Genton14}
Genton, M.~G. and Kleiber, W. (2014).
\newblock Cross-covariance functions for multivariate geostatistics.
\newblock {\em Statistical Science}, 30(2):147--163.

\bibitem[Glahn and Lowry, 1972]{Glahn72}
Glahn, H.~R. and Lowry, D.~A. (1972).
\newblock The use of model output statistics (mos) in objective weather
  forecasting.
\newblock {\em Journal of Applied Meteorology}, 11(8):1203--1211.

\bibitem[Gneiting et~al., 2006]{Gneiting06}
Gneiting, T., Larson, K., Westrick, K., Genton, M.~G., and Aldrich, E. (2006).
\newblock Calibrated probabilistic forecasting at the stateline wind energy
  center: The regime-switching space--time method.
\newblock {\em Journal of the American Statistical Association},
  101(475):968--979.

\bibitem[Gneiting et~al., 2005]{Gneiting05}
Gneiting, T., Raftery, A.~E., Westveld~III, A.~H., and Goldman, T. (2005).
\newblock Calibrated probabilistic forecasting using ensemble model output
  statistics and minimum {CRPS} estimation.
\newblock {\em Monthly Weather Review}, 133(5):1098--1118.

\bibitem[Gneiting et~al., 2008]{Gneiting08}
Gneiting, T., Stanberry, L.~I., Grimit, E.~P., Held, L., and Johnson, N.~A.
  (2008).
\newblock Assessing probabilistic forecasts of multivariate quantities, with an
  application to ensemble predictions of surface winds.
\newblock {\em Test}, 17(2):211--235.

\bibitem[Hamill, 2001]{Hamill01}
Hamill, T.~M. (2001).
\newblock Interpretation of rank histograms for verifying ensemble forecasts.
\newblock {\em Monthly Weather Review}, 129(3):550--560.

\bibitem[Hering and Genton, 2010]{Hering10}
Hering, A.~S. and Genton, M.~G. (2010).
\newblock Powering up with space-time wind forecasting.
\newblock {\em Journal of the American Statistical Association},
  105(489):92--104.

\bibitem[Kang et~al., 2012]{Kang12}
Kang, E.~L., Cressie, N., and Sain, S.~R. (2012).
\newblock Combining outputs from the {N}orth {A}merican regional climate change
  assessment program by using a {B}ayesian hierarchical model.
\newblock {\em Journal of the Royal Statistical Society: Series C (Applied
  Statistics)}, 61(2):291--313.

\bibitem[Lerch and Thorarinsdottir, 2013]{Lerch13}
Lerch, S. and Thorarinsdottir, T.~L. (2013).
\newblock Comparison of nonhomogeneous regression models for probabilistic wind
  speed forecasting.
\newblock {\em arXiv preprint arXiv:1305.2026}.

\bibitem[Li et~al., 2015]{Li15}
Li, N., Uckun, C., Constantinescu, E., Birge, J., Hedman, K., and Botterud, A.
  (2015).
\newblock Flexible operation of batteries in power system scheduling with
  renewable energy.
\newblock {\em IEEE Transactions on Sustainable Energy, in print}.

\bibitem[Majda and Wang, 2006]{Majda06}
Majda, A. and Wang, X. (2006).
\newblock {\em Nonlinear dynamics and statistical theories for basic
  geophysical flows}.
\newblock Cambridge University Press.

\bibitem[Palmer, 2014]{Palmer14}
Palmer, T. (2014).
\newblock More reliable forecasts with less precise computations: a fast-track
  route to cloud-resolved weather and climate simulators?
\newblock {\em Philosophical Transactions of the Royal Society of London A:
  Mathematical, Physical and Engineering Sciences}, 372(2018):20130391.

\bibitem[Papavasiliou et~al., 2015]{Papavasiliou15}
Papavasiliou, A., Oren, S.~S., and Rountree, B. (2015).
\newblock Applying high performance computing to transmission-constrained
  stochastic unit commitment for renewable energy integration.
\newblock {\em IEEE Transactions on Power Systems}, 30(3):1109--1120.

\bibitem[Pinson, 2013]{Pinson13}
Pinson, P. (2013).
\newblock Wind energy: {F}orecasting challenges for its operational management.
\newblock {\em Statistical Science}, 28(4):564--585.

\bibitem[Pinson et~al., 2008]{Pinson08}
Pinson, P., Christensen, L. E.~A., Madsen, H., S\/orensen, P.~E., Donovan,
  M.~H., and E., J.~L. (2008).
\newblock Regime-switching modelling of the fluctuations of offshore wind
  generation.
\newblock {\em Journal of Wind Engineering and Industrial Aerodynamics},
  96(12):2327--2347.

\bibitem[Pinson and Girard, 2012]{Pinson12}
Pinson, P. and Girard, R. (2012).
\newblock Evaluating the quality of scenarios of short-term wind power
  generation.
\newblock {\em Applied Energy}, 96:12--20.

\bibitem[Pinson et~al., 2009]{Pinson09}
Pinson, P., Madsen, H., Nielsen, H., Papaefthymiou, G., and Kl{\"o}ckl, B.
  (2009).
\newblock From probabilistic forecasts to statistical scenarios of short-term
  wind power production.
\newblock {\em Wind Energy}, 12(1):51--62.

\bibitem[Pinson and Tastu, 2013]{Pinson13b}
Pinson, P. and Tastu, J. (2013).
\newblock Discrimination ability of the energy score.
\newblock Technical report, Technical University of Denmark.

\bibitem[Raftery et~al., 2005]{Raftery05}
Raftery, A.~E., Gneiting, T., Balabdaoui, F., and Polakowski, M. (2005).
\newblock Using {B}ayesian model averaging to calibrate forecast ensembles.
\newblock {\em Monthly Weather Review}, 133(5):1155--1174.

\bibitem[Royle and Berliner, 1999]{Royle99a}
Royle, J. and Berliner, L. (1999).
\newblock A hierarchical approach to multivariate spatial modeling and
  prediction.
\newblock {\em Journal of Agricultural, Biological, and Environmental
  Statistics}, pages 29--56.

\bibitem[Royle et~al., 1999]{Royle99b}
Royle, J., Berliner, L., Wikle, C., and Milliff, R. (1999).
\newblock A hierarchical spatial model for constructing wind fields from
  scatterometer data in the {L}abrador {S}ea.
\newblock In {\em Case Studies in Bayesian Statistics}, pages 367--382.
  Springer.

\bibitem[Schefzik et~al., 2013]{Schefzik13}
Schefzik, R., Thorarinsdottir, T.~L., and Gneiting, T. (2013).
\newblock Uncertainty quantification in complex simulation models using
  ensemble copula coupling.
\newblock {\em Statistical Science}, 28(4):616--640.

\bibitem[Scheuerer and Hamill, 2015]{Scheuerer15b}
Scheuerer, M. and Hamill, T.~M. (2015).
\newblock Variogram-based proper scoring rules for probabilistic forecasts of
  multivariate quantities.
\newblock {\em Monthly Weather Review}, 143(4):1321--1334.

\bibitem[Scheuerer and M{\"o}ller, 2015]{Scheuerer15}
Scheuerer, M. and M{\"o}ller, D. (2015).
\newblock Probabilistic wind speed forecasting on a grid based on ensemble
  model output statistics.
\newblock {\em The Annals of Applied Statistics}, 9(3):1328--1349.

\bibitem[Schuhen et~al., 2012]{Schuhen12}
Schuhen, N., Thorarinsdottir, T.~L., and Gneiting, T. (2012).
\newblock Ensemble model output statistics for wind vectors.
\newblock {\em Monthly Weather Review}, 140:3204--3219.

\bibitem[Shumway and Stoffer, 2010]{Shumway10}
Shumway, R. and Stoffer, D. (2010).
\newblock {\em Time series analysis and its applications: with {R} examples}.
\newblock Springer Science \& Business Media.

\bibitem[Skamarock et~al., 2008]{Skamarock08}
Skamarock, W., Klemp, J., Dudhia, J., Gill, D., Barker, D., Duda, M., Huang,
  X.-Y., Wang, W., and Powers, J. (2008).
\newblock A description of the {A}dvanced {R}esearch {WRF} version 3.
\newblock Technical Report Tech Notes-475+ STR, NCAR.

\bibitem[Sloughter et~al., 2010]{Sloughter10}
Sloughter, J. M.~L., Gneiting, T., and Raftery, A.~E. (2010).
\newblock Probabilistic wind speed forecasting using ensembles and {B}ayesian
  model averaging.
\newblock {\em Journal of the {A}merican {S}tatistical {A}ssociation},
  105(489):25--35.

\bibitem[Sloughter et~al., 2013]{Sloughter13}
Sloughter, J. M.~L., Gneiting, T., and Raftery, A.~E. (2013).
\newblock Probabilistic wind vector forecasting using ensembles and {B}ayesian
  model averaging.
\newblock {\em Monthly Weather Review}, 141(6):2107--2119.

\bibitem[Smith and Hansen, 2004]{Smith04}
Smith, L.~A. and Hansen, J.~A. (2004).
\newblock Extending the limits of ensemble forecast verification with the
  minimum spanning tree.
\newblock {\em Monthly Weather Review}, 132(6):1522--1528.

\bibitem[Stein et~al., 2012]{Stein12b}
Stein, M., Chen, J., and Anitescu, M. (2012).
\newblock Difference filter preconditioning for large covariance matrices.
\newblock {\em SIAM Journal on Matrix Analysis and Applications}, 33(1):52--72.

\bibitem[Stein, 2012]{Stein12}
Stein, M.~L. (2012).
\newblock {\em Interpolation of spatial data: some theory for kriging}.
\newblock Springer Science \& Business Media.

\bibitem[Thorarinsdottir and Gneiting, 2010]{Thorarinsdottir10}
Thorarinsdottir, T.~L. and Gneiting, T. (2010).
\newblock Probabilistic forecasts of wind speed: ensemble model output
  statistics by using heteroscedastic censored regression.
\newblock {\em Journal of the Royal Statistical Society: Series A (Statistics
  in Society)}, 173(2):371--388.

\bibitem[Thorarinsdottir and Johnson, 2012]{Thorarinsdottir12}
Thorarinsdottir, T.~L. and Johnson, M.~S. (2012).
\newblock Probabilistic wind gust forecasting using non-homogeneous {G}aussian
  regression.
\newblock {\em Monthly Weather Review}, 140(3):889--897.

\bibitem[Thorarinsdottir et~al., 2014]{Thorarinsdottir14}
Thorarinsdottir, T.~L., Scheuerer, M., and Heinz, C. (2014).
\newblock Assessing the calibration of high-dimensional ensemble forecasts
  using rank histograms.
\newblock {\em Journal of Computational and Graphical Statistics},
  (just-accepted):00--00.

\end{thebibliography}

\begin{flushright}
 \scriptsize \framebox{\parbox{3.2in}{
The submitted manuscript has been created by UChicago Argonne, LLC, Operator of Argonne National Laboratory (Argonne). Argonne, a U.S. Department of Energy Office of Science laboratory, is operated under Contract No. DE-AC02-06CH11357. The U.S. Government retains for itself, and others acting on its behalf, a paid-up nonexclusive, irrevocable worldwide license in said article to reproduce, prepare derivative works, distribute copies to the public, and perform publicly and display publicly, by or on behalf of the Government.  The Department of Energy will provide public access to these results of federally sponsored research in accordance with the DOE Public Access Plan. \url{http://energy.gov/downloads/doe-public-access-plan}. }}
\end{flushright}

\end{document}